\documentclass{aims} 
\usepackage{amsmath}
\usepackage{paralist}
\usepackage[misc]{ifsym}
\usepackage{epsfig} 
\usepackage{epstopdf} 
\usepackage[colorlinks=true]{hyperref}
\hypersetup{urlcolor=blue, citecolor=red}
\allowdisplaybreaks

 \def\currentvolume{X}
  \def\currentissue{X}
   \def\currentyear{200X}
    \def\currentmonth{XX}
     \def\ppages{X-XX}
      \def\DOI{10.3934/xx.xxxxxxx}

\theoremstyle{definition}

\usepackage{subfigure}

\usepackage{bm}
\usepackage{mathrsfs}
\usepackage{amsmath}
\usepackage{amsfonts}
\usepackage{amsthm}
\usepackage[ruled]{algorithm} 
\usepackage[noend]{algpseudocode} 
\usepackage{color}
\usepackage{setspace}
\usepackage{float}
\usepackage{import}
\usepackage{url}
\usepackage{multicol}
\usepackage{multirow}
\usepackage{subfigure}

\usepackage{tikz}

\newcommand{\real}{\mathbb{R}}

\newcommand{\mcD}{\mathcal{D}}

\newcommand{\mcG}{\mathcal{G}}

\newcommand{\mcU}{\mathcal{U}}

\newcommand{\mcN}{\mathcal{N}}

\usepackage{xcolor}

\def\omg{{\Omega}}

\def \fb{\bm{f}}

\def \ub{\bm{u}}

\def \vb{\bm{v}}
\def \xb{\bm{x}}

\def \hb{\bm{h}}

\def \nb{\bm{n}}

\def \cb{\bm{c}}

\def \xib{{\boldsymbol\xi}}

\newcommand{\vertii}[1]{{\left\vert\left\vert #1
    \right\vert\right\vert}}

\usepackage{lineno}

\title[Deep Neural Operator Enabled Digital Twin Modeling for AM]
{Deep Neural Operator Enabled Digital Twin Modeling for Additive Manufacturing} 

\author[Ning Liu et al.]{}

\subjclass{Primary: 68T07, 93E35, 74M05; Secondary: 35S30, 41A35.}
\keywords{Digital Twin, Neural Operators, Closed-Loop Feedback Control, Laser Powder Bed Fusion, Physics-Informed Modeling, Defect Characterization.}

\begin{document}
\maketitle

\centerline{\scshape
Ning Liu$^{{\href{mailto:nliu@gem-innovation.com}{\textrm{\Letter}}}1*}$,
Xuxiao Li$^{{\href{mailto:xli@gem-innovation.com}{\textrm{\Letter}}}1}$,
Manoj R. Rajanna$^{{\href{mailto:mrajanna@gem-innovation.com}{\textrm{\Letter}}}1}$,
Edward W. Reutzel$^{{\href{mailto:ewr101@arl.psu.edu}{\textrm{\Letter}}}2}$,
}

\centerline{\scshape
Brady Sawyer$^{{\href{mailto:bas6156@arl.psu.edu}{\textrm{\Letter}}}2}$,
Prahalada Rao$^{{\href{mailto:prahalad@vt.edu}{\textrm{\Letter}}}3}$,
Jim Lua$^{{\href{mailto:jlua@gem-innovation.com}{\textrm{\Letter}}}1}$,
Nam Phan$^{4}$,
and Yue Yu$^{{\href{mailto:yuy214@lehigh.edu}{\textrm{\Letter}}}5*}$}

\medskip

{\footnotesize
 \centerline{$^1$Global Engineering and Materials, Inc., Princeton, NJ 08540, USA}
} 

\medskip

{\footnotesize
 \centerline{$^2$Applied Research Laboratory, The Pennsylvania State University, University Park, PA 16802, USA}
}



\medskip

{\footnotesize
 \centerline{$^3$Grado Department of Industrial and Systems Engineering, Virginia Tech, Blacksburg, VA 24061, USA}
}

\medskip

{\footnotesize
 \centerline{$^4$Structures Division, Naval Air Systems Command (NAVAIR), Patuxent River, MD 20670, USA}
}

\medskip

{\footnotesize
 \centerline{$^5$Department of Mathematics, Lehigh University, Bethlehem, PA 18015, USA}
}

\medskip

{\footnotesize
 \centerline{$^*$Corresponding Author(s)}
}

\bigskip

 \centerline{(Communicated by Handling Editor)}


\begin{abstract}
A digital twin (DT), with the components of a physics-based model, a data-driven model, and a machine learning (ML) enabled efficient surrogate, behaves as a virtual twin of the real-world physical process. In terms of Laser Powder Bed Fusion (L-PBF) based additive manufacturing (AM), a DT can predict the current and future states of the melt pool and the resulting defects corresponding to the input laser parameters, evolve itself by assimilating in-situ sensor data, and optimize the laser parameters to mitigate defect formation. In this paper, we present a deep neural operator enabled computational framework of the DT for closed-loop feedback control of the L-PBF process. This is accomplished by building a high-fidelity computational model to accurately represent the melt pool states, an efficient surrogate model to approximate the melt pool solution field, followed by an physics-based procedure to extract information from the computed melt pool simulation that can further be correlated to the defect quantities of interest (e.g., surface roughness). In particular, we leverage the data generated from the high-fidelity physics-based model and train a series of Fourier neural operator (FNO) based ML models to effectively learn the relation between the input laser parameters and the corresponding full temperature field of the melt pool. The learned FNO models are able to make fast inference to assist in real-time monitoring and control. Subsequently, a set of physics-informed variables such as the melt pool dimensions and the peak temperature can be extracted to compute the resulting defects. An optimization algorithm is then exercised to control laser input and minimize defects. On the other hand, the constructed DT can also evolve with the physical twin via offline finetuning and online material calibration. For instance, the probabilistic distribution of laser absorptivity can be updated to match the real-time captured thermal image data. Finally, a probabilistic framework is adopted for uncertainty quantification. The developed DT is envisioned to guide the AM process and facilitate high-quality manufacturing in L-PBF-based metal AM.

\end{abstract}


\tableofcontents

\section{Introduction}
Recent advances in additive manufacturing (AM) technologies \cite{frazier2014metal,gibson2021additive,abdulhameed2019additive} have received considerable attention from both the academic and industrial communities. In particular, the Laser Powder Bed Fusion (L-PBF) in AM \cite{king2015laser,yadroitsev2021fundamentals} offers a wide variety of advantages, including reduced material waste, improved turnaround time, enhanced flexibility in quickly iterating on a design perspective, and high-quality printing applicable to a large amount of materials. However, the quality of the printed parts in L-PBF is highly reliant on the prescribed process parameters that control the microstructure formation. Poor process conditions can lead to undesired formation of defects such as lack-of-fusion porosity \cite{tang2017prediction,promoppatum2022quantification}, surface roughness \cite{snyder2020understanding,fox2016effect}, and keyhole \cite{zhao2020critical,kouraytem2019effect}, which are often attributed to irregular formations of the melt pool. Although high-fidelity physics-based simulations can provide insights into the root cause of such unwanted phenomena, it is computationally prohibitive to provide real-time advice for one thing and, for another, they cannot closely capture the evolution of the printing process. The demand for printing high-quality parts calls for an intelligent agent with which one can actively monitor and control the printing process. To this end, digital twins \cite{kapteyn2021probabilistic,gunasegaram2021towards,mukherjee2019digital} are invented to bridge the divide between the physical experiments and computational modeling and closely steer the manufacturing process.

\begin{figure}[!t]\centering
\includegraphics[width=1.0\textwidth]{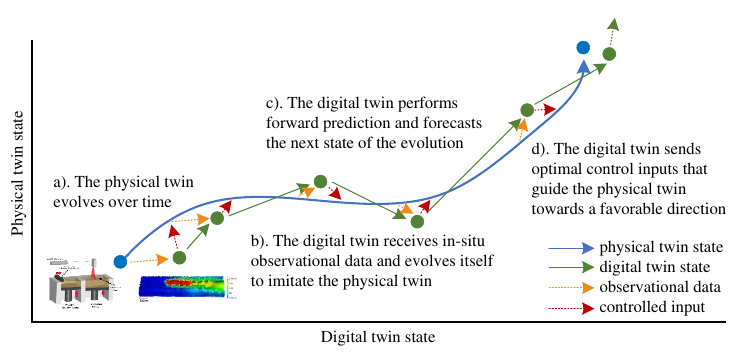}
  \caption{Demonstration of the digital twin concept in additive manufacturing.}\label{fig:dt_concept}
\end{figure}

A digital twin (DT) \cite{kapteyn2021probabilistic,kharazmi2021data}, with the components of a physics-based model, a data-driven model, and a machine learning (ML) enabled surrogate model, behaves as a virtual twin of the physical AM process. The general concept of the DT is demonstrated in Figure~\ref{fig:dt_concept}. A simple way to construct an initial DT is to train an offline ML model using a robust database established by a combination of the physics-based modeling results and available experimental data, such that the DT closely mimics the behavior of the physical twin (PT). The DT takes as input the intrinsic parameters such as material properties and controllable parameters such as the laser power and scan speed, and outputs user-cared information such as the full-scale temperature field of the melt pool, defects, and microstructure parameters. During experiment, the state of the PT changes over time, which is captured by in-situ monitored observational data such as thermal images. The DT can then assimilate the real-time data, self-evolve to improve prediction accuracy, predict the next state(s) of the evolution, and intelligently control the process parameters to achieve a desired setpoint, such as a preferred microstructure or reduced defects.

As one of the essential building blocks, high-fidelity physics-based models serve as the foundation towards constructing a DT. In the literature, there exist a wide variety of physics-based process models with various fidelities, resolutions, and computation cost \cite{li2021three,moges2021hybrid,queva2020numerical,khorasani2022comprehensive,yeung2020meltpool}. For example, the most complex computational model can resolve keyhole, melt pool, vapor plume, and spatter \cite{li2021quantitative}, while a simplified model only considers heat conduction and captures melt pool formation \cite{heigel2015thermo}. Although high-fidelity computational models are effective in capturing the melt pool geometries and accurately predicting both the thermal and velocity fields, the associated high computational cost \cite{strano2013surface} hinders their application for massive data generation in a data-driven analysis setting. In the current work, we choose to employ a relatively simplified model for efficient parametric sweeping of process parameters for ML data generation.

On the other hand, several ML algorithms have been utilized towards defect mitigation \cite{liu2023review,sing2021perspectives,okaro2019automatic,du2021physics,wang2022process,liao2024deep}. In \cite{larsen2022deep}, a semi-supervised variational recurrent neural network (VRNN) is proposed to model the offset between the predicted and observed states and detect anomaly directly from high-speed image data. This method takes into account the time-series effect between images and explores the correlation between observational data and laser input to reduce false detections. The work of Jin et al. \cite{tan2019encoder} reports an unsupervised encoder-decoder approach to detect anomaly and predict the next step of the printing process. Worth noting is the work done by Li et al. \cite{li2023statistical}, where a statistical parameterized physics-based DT is built for closed-loop control of L-PBF. This work introduces a reduced-order stochastic calibration process to enable the statistical prediction of the melt pool status and the resulting defects. Nevertheless, their employed autoregressive network only aims to predict the width and depth of the melt pool and cannot be used as a high-fidelity solution surrogate. As a result, the generalizability of the model is questionable. The current study takes a different approach and investigates learning a high-fidelity physics surrogate of the L-PBF process with neural operators \cite{you2022learning,li2020fourier,liu2023domain,liu2023clawno,jafarzadeh2024peridynamic} as the backbone, which aim to learn the hidden physics of the underlying process and feature resolution-independence and robust generalizability to unseen input instances \cite{liu2023ino}.

\begin{figure}[!t]\centering
\includegraphics[width=1.0\textwidth]{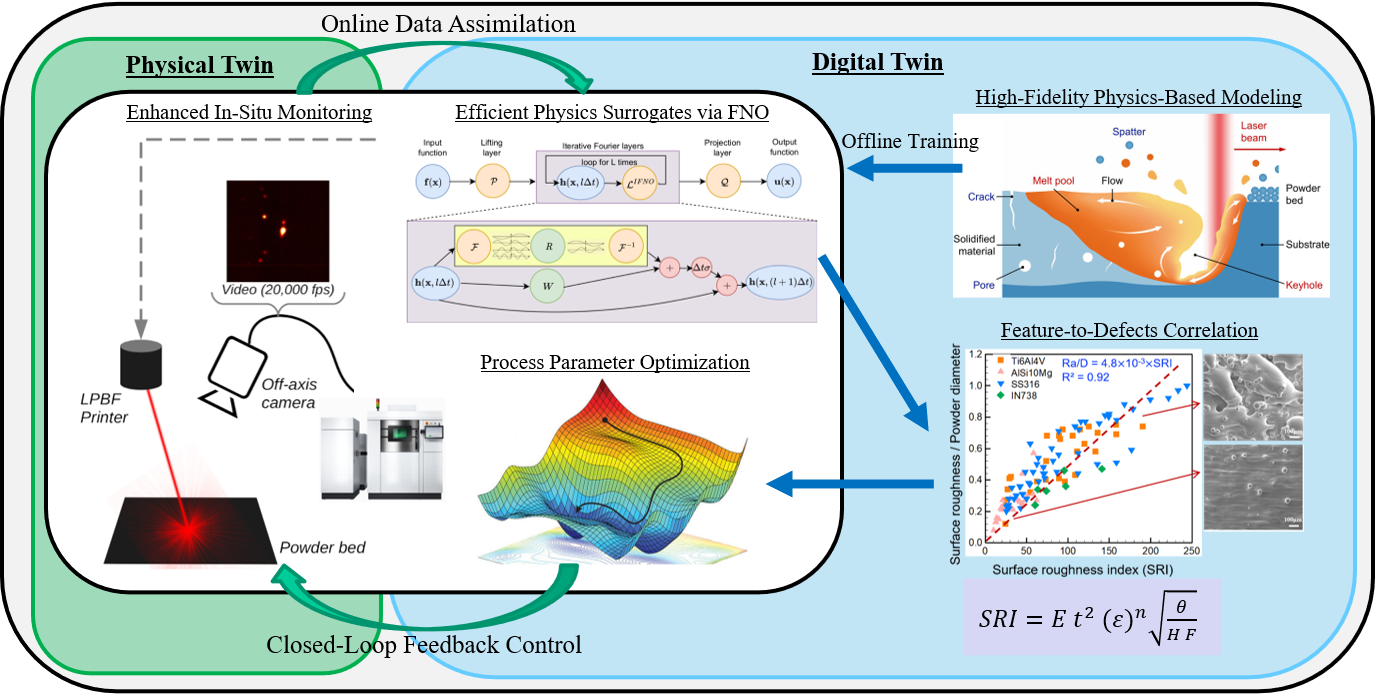}
  \caption{An overview of the proposed DT framework.}\label{fig:dt_overview}
\end{figure}

In this context, we propose an integrated DT framework for closed-loop feedback control of the L-PBF process. An overview of the proposal DT framework is illustrated in Figure~\ref{fig:dt_overview}. We start the DT construction process by establishing a high-fidelity database from physics-based simulations. A series of FNOs are then trained offline to accurately approximate the physics-based solution field. Next, an effective correlation between the melt pool states and the resulting defects of interest is established, with which the process parameter optimization is made possible via designing an objective function to minimize defects and arrive at the optimal process parameters leveraging automatic differentiation in ML. The DT then informs the PT of the computed optimal conditions while in the same time assimilates the in-situ monitored data and evolves itself to closely represent the current state of the PT. The DT is also finetuned using newly collected data after the experiment is finished.

This paper is organized as follows. In Section~\ref{sec:physics_modeling}, a physics-based modeling approach of the L-PBF process is introduced, along with a data-driven method for defect characterization. The main concept of neural operator enabled physics surrogate modeling is presented in Section~\ref{sec:no}, where an optimization algorithm is designed to compute optimal process parameters and minimize defects. We present in Section~\ref{sec:integrated_dt} the integrated probabilistic DT framework with uncertainty quantification, followed by a virtual demonstration of the proposed DT in reducing surface roughness as an example in Section~\ref{sec:virtual_demo}. The obtained results comparing defects with and without DT control are discussed thoroughly in Section~\ref{sec:discussion} and conclusions are drawn in Section~\ref{sec:conclusion}. To provide a condensed overview of the proposed algorithm, a summary of the key ingredients and steps can be found in Algorithm \ref{alg:ML}.

\begin{algorithm}
\caption{Key steps for deep FNO enabled DT.}\label{alg:ML}
\begin{algorithmic}[1]
\State \textbf{Offline training phase:}
\State Read in input parameter sets $\xib:=(P,V,T_{sub},\alpha)$ and the corresponding temperature fields $T(x,y,z)$: $\mcD_{tr}:=\{\xib_i,T_i(x,y,z)\}_{i=1}^{N_{tr}}$ for training, $\mcD_{val}:=\{\xib^{val}_i,T^{val}_i(x,y,z)\}_{i=1}^{N_{val}}$ for validation.
\State Interpolate the simulated temperature fields with spline functions, to the two predefined grids for the $x-y$ and $x-z$ planes and form the datasets for two FNOs:
$$\mcD_{x-y,tr}:=\{\xib_i,T_{i,x-y}(x,y)|_{(x,y)\in\chi_{x-y}}\}_{i=1}^{N_{tr}},$$
$$\mcD_{x-y,val}:=\{\xib^{val}_i,T^{val}_{i,x-y}(x,y)|_{(x,y)\in\chi_{x-y}}\}_{i=1}^{N_{val}},$$
$$\mcD_{x-z,tr}:=\{\xib_i,T_{i,x-z}(x,z)|_{(x,z)\in\chi_{x-z}}\}_{i=1}^{N_{tr}},$$
$$\mcD_{x-z,val}:=\{\xib^{val}_i,T^{val}_{i,x-z}(x,z)|_{(x,z)\in\chi_{x-z}}\}_{i=1}^{N_{val}}.$$
\State Construct the FNO models, $\hat{\mcG}_1[\xib;\theta_1]$ and $\hat{\mcG}_2[\xib;\theta_2]$, to match the temperature field on the $x-y$ and $x-z$ planes, respectively:
$$\theta_1=\underset{\theta_1\in\Theta}{\text{argmin}}\sum_{i=1}^{N_{tr}}\sum_{\xb_j\in \chi_{x-y}}\vertii{\hat{\mcG}_1[\xib_i;\theta_1](\xb_j)-T_i(\xb_j)}^2,$$
$$\theta_2=\underset{\theta_2\in\Theta}{\text{argmin}}\sum_{i=1}^{N_{tr}}\sum_{\xb_j\in \chi_{x-z}}\vertii{\hat{\mcG}_2[\xib_i;\theta_1](\xb_j)-T_i(\xb_j)}^2.$$
\State \textbf{Online data assimilation phase:}
\State Extract melt pool dimensions from real-time captured thermal image data and compute their distributions.
\State Initialize the estimated mean $\mu_\alpha$ and standard deviation $\sigma_\alpha$ for the laser absorptivity $\alpha$, which is taken as the quantity with uncertainty in this work.
\For{$ep = 1: epoch_{max}$}
\State Sample $\alpha\sim\mcN[\mu_\alpha,\sigma^2_\alpha]$ for $S$ times.
\For{$s = 1: S$}
\State Run a forward pass of the two FNO models based on the $s-$th sample: $\xib_s=[P,V,T_{sub},\alpha_s]$, so as to estimate the corresponding temperature field $T$.
\State Estimate the melt pool length following \eqref{eq:Tlength}.
\EndFor
\State Estimate the mean and standard deviation of melt pool length following \eqref{eq:meanL}-\eqref{eq:stdL}.
\State Calculate the Kullback–Leibler divergence (see \eqref{eq:calialpha}) between observed and predicted distributions of melt pool dimensions.
\State Update $\mu_\alpha$, $\sigma_\alpha$ to minimize \eqref{eq:calialpha}, with the Adam optimizer.
\EndFor
\State With the inferred distribution of laser absorptivity, estimate statistical moments of the corresponding solutions and other quantities of interests.
\State \textbf{Feedback control phase:}
\State Initialize the estimated laser power $P$ and scan speed $V$, which are taken as the controllable quantities in this work.
\For{$ep = 1: epoch_{max}$}
\State Run a forward pass of the two FNO models, so as to estimate the corresponding temperature field $T$.
\State Extract melt pool dimensions and peak temperature and compute the surface roughness index (SRI), following \eqref{eq:Tpeak}-\eqref{eq:Tratio}.
\State Estimate the corresponding surface roughness following \eqref{eq:ra}.
\State Update $P$, $V$ to minimize \eqref{eqn:optPV}, with the Adam optimizer.
\EndFor
\end{algorithmic}
\end{algorithm}

\section{Physics-Based Modeling}\label{sec:physics_modeling}

\subsection{Physics-Based Process Simulation}\label{sec:physics_model}

We aim to employ a proper physics-based model to simulate the critical field functions such as the 3D temperature field associated with defects formation. The physics-based model is used to create a database, based on which the ML models can be trained to rapidly predict field functions and defects. In the literature, there exist a large variety of process models with different fidelities, resolutions, and computation cost. For example, the most complex model can resolve keyhole, melt pool, vapor plume, and spatter \cite{li2021quantitative}, while a simplified model only considers heat conduction and captures melt pool formation \cite{heigel2015thermo}. Here, we adopt a simplified model for sake of computational time, where the governing physics is heat conduction and the effects of fluid flow, keyhole, and vapor plume are neglected for now. We aim to build a general workflow and the synergy between physics-based and ML models for creation of a DT. In particular, considering a three dimensional computational domain $\Omega:= [0,10000]\times[-1000,1000]\times[0,1000]$ $(\mu m^3)$, the governing equation of temperature $T$ can be written as:
\begin{align}
&\frac{\partial \left( \rho h\left(T(x,y,z,t)\right)\right)}{\partial t} = k \nabla^2 T(x,y,z,t) + \dot{S}(x,y,z,t), \text{ for }(x,y,z,t)\in \Omega\times[0,T],\label{eq:energy_conserv}\\
&T(x,y,z,0)=T_{sub}, \text{ for }(x,y,z)\in \Omega,\\
&\dfrac{\partial T(x,y,z,t)}{\partial \nb}=0, \text{ if }(x,y,z)\in \partial\Omega \text{ and }z\neq 1000,\\
&-k\dfrac{\partial T(x,y,z,t)}{\partial z}=h_0 (T_\infty - T(x,y,z,t)), \text{ if }z= 1000,
\end{align}
where $\rho$ represents density, $h$ is the specific enthalpy, $k$ denotes thermal conductivity which is taken as the solid thermal conductivity $k_s$ or the liquid thermal conductivity $k_l$, and $T_{sub}$ is the substrate temperature which is also set as the initial condition. The top surface has a convective heat transfer coefficieint of $h_0=1 W/(m^2 \cdot K)$, and the ambient temperature is set to $T_\infty=300K$. $\dot{S}$ describes the heat source term due to laser heating, which can be further expanded as:
\begin{equation}\label{eq:heat_source}
\dot{S}(x,y,z) = \dfrac{\alpha P}{\pi r^2_b t_h} \exp\left(-\dfrac{2r^2(x,y,z)}{r^2_b}\right) \text{ ,}
\end{equation}
where $P$ is the laser power, $t_h$ is layer thickness, $\alpha$ denotes absorptivity, and $r_b$ is the laser radius. $r(x,y,z):=\sqrt{(x-x_l)^2+(y-y_l)^2+(z-z_l)^2}$ is the radial distance of each spatial point, $(x,y,z)$, to the laser center, $(x_l,y_l,z_l)$. In this work, we assume that the laser center moves under a fixed scan speed, $V$, along the direction of $x$. Note that $\dot{S}$ is only in the powder layer and is deactivated elsewhere. That means, we have $\dot{S}(x,y,z)=0$ for $z<1000$.

\begin{figure}[!t]\centering
\includegraphics[width=0.7\textwidth]{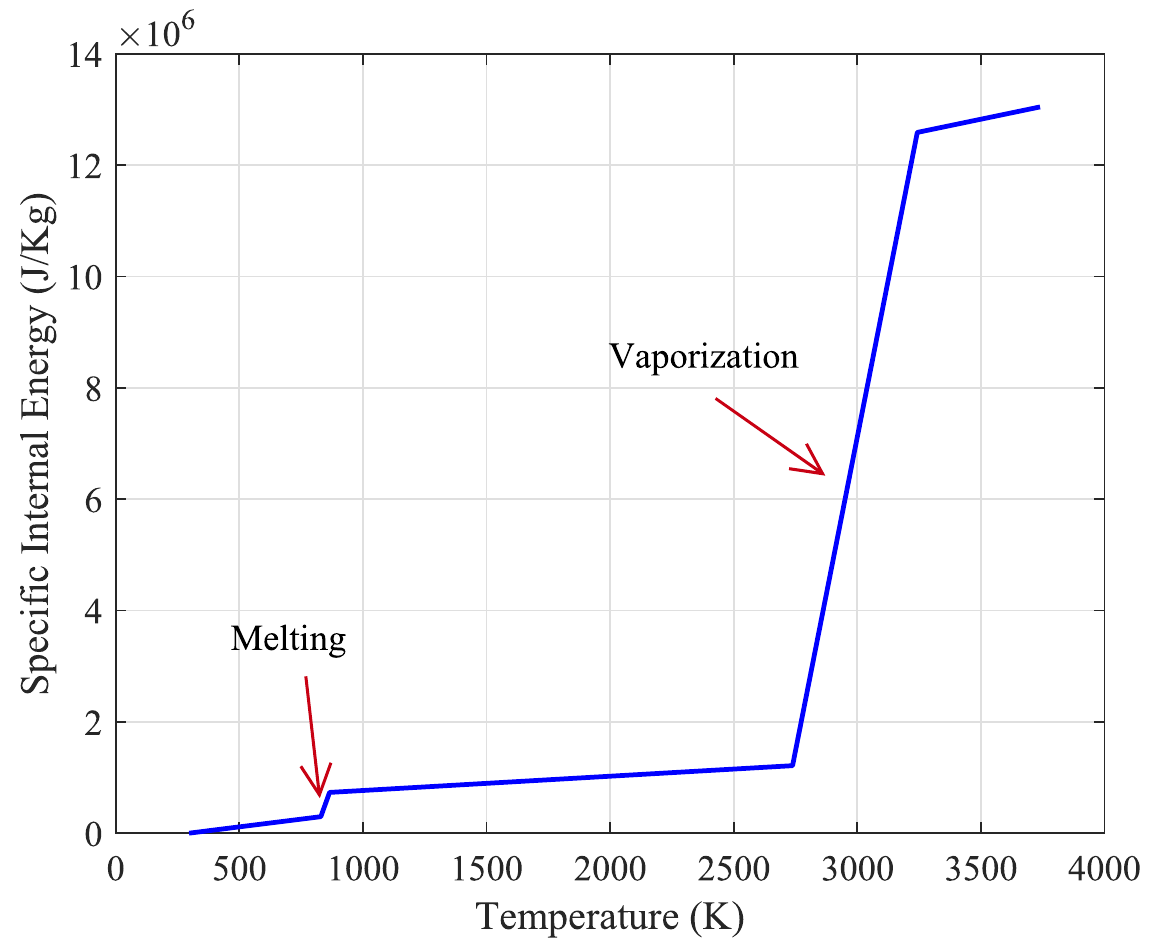}
  \caption{Enthalpy formulation for melting and vaporization.}\label{fig:enthalpy_formulation}
\end{figure}

The enthalpy formulation in \eqref{eq:energy_conserv} is used to consider melting and solidification. To eliminate non-physical large temperature, the vaporization is considered by formulating an enthalpy jump at the boiling temperature, as shown in Figure~\ref{fig:enthalpy_formulation}. The material properties used in the simulation for AlSi10Mg are listed in Table~\ref{tab:mat_prop_AlSi10Mg}. All physics-based simulations are conducted using an adaptive finite element analysis model based on STAR-CCM+ \cite{piasecka2022experimental}. For each input combination $\xib:=(P,V,T_{sub},\alpha)$, we numerically solve \eqref{eq:energy_conserv} until its solution, $T$, arrives at a steady state. With a slight abuse of notation, in the following we refer to this steady-state solution as $T(x,y,z)$. The most refined mesh has a size of 3.125 $\mu m$ near the molten pool region, and the mesh size is coarsened to 12.5 $\mu m$ gradually. With this meshing scheme, it takes approximately 20 min to finish one simulation using 48 CPU cores (average clock speed is 1.4GHz), which is non-feasible for the real-time monitoring and parameter calibration. Therefore, ML models are imperative to enable rapid prediction and two-way interaction between physical and digital twins, as will be elaborated in Section \ref{sec:no}.

\begin{table}[!h]
    \caption{Material properties for AlSi10Mg.}
    \label{tab:mat_prop_AlSi10Mg}
    \centering
    {\small \centering
    \begin{tabular}{ll}
    \hline
         Density $\rho$ & 2670 $kg/m^3$\\
         Solid specific heat & 546 $J/(kg \cdot K)$\\
         Liquid specific heat & 632 $J/(kg \cdot K)$\\
         Latent heat of melting & $4.23 \times 10^5 J/kg$\\
         Latent heat of vaporization & $1.14 \times 10^7 J/kg$\\
         Solid thermal conductivity $k_s$ & 113 $W/(m \cdot K)$\\
         Liquid thermal conductivity $k_l$ & 133 $W/(m \cdot K)$\\
         Solidus temperature $T_s$ & 831 $K$\\
         Liquidus temperature $T_l$ & 867 $K$\\
         Boiling temperature $T_b$ & 2740 $K$\\
         Absorptivity $\alpha$ & 0.3 \\
         \hline
    \end{tabular}}
\vskip -0.1in
\end{table}

\subsection{Data-Driven Defect Characterization}\label{sec:defect_characterization}

The surface roughness model is based on the Buckingham Pi theorem \cite{du2022high}. Specifically, the surface roughness index (SRI) is calculated based on the estimated temperature field $T$, and it is then used to indicate the surface roughness (Ra). The calculation of SRI from $T$ can be written in the following form:
\begin{equation}\label{eq:sri}
SRI[T] = E \hat{L}^2\epsilon[T]^{0.25} \sqrt{\frac{\beta}{\hat{H} F[T]}} \text{ ,}
\end{equation}
where $E$ is the letant heat of melting, $\hat{L}$ is a characteristic length scale taken to be 50 $\mu m$, $\epsilon[T]$ is the ratio between melt pool length $L[T]$ and width $W[T]$, 
$\beta$ is the contact angle which is taken to be 1.1, $\hat{H}$ is a characteristic energy density taken to be 150 $J/m^2$, and $F$ is the Marangoni force which is dependent on $T$:
\begin{equation}\label{eq:Marangoni_force}
F[T] = \gamma_T(T_\text{peak}[T] - T_\text{s}) \times \frac{\pi L[T]}{2} \text{ .}
\end{equation}
Here, $\gamma_T$ is the temperature derivative of surface tension (taken to be 0.00035 $N/(m \cdot K)$), $T_\text{peak}[T]$ is the peak temperature in the melt pool, and $T_\text{s}$ is the solidus temperature. With a given temperature field, $T(x,y,z)$, we can calculate the peak temperature $T_\text{peak}[T]$, the melt pool length $L[T]$, the melt pool width $W[T]$, and their ratio $\epsilon[T]$ via:
\begin{align}
&T_\text{peak}[T]=\max_{x,y,z} T(x,y,z),\\
&L[T]=\max_{y,z} \int_{x=-5000}^{5000}\mathbf{1}(T(x,y,z)-T_s) dx,\\
&W[T]=\max_{x,z} \int_{y=-1000}^{1000}\mathbf{1}(T(x,y,z)-T_s) dy,\\
&\epsilon[T]=L[T]/W[T].
\end{align}
Here, $\mathbf{1}$ is a characteristic function taken as:
\begin{equation}
\mathbf{1}(f(x,y,z)):=\left\{\begin{array}{rl}
1,& \text{ if }f(x,y,z)>0;\\
0,& \text{ else }.
\end{array}\right.
\end{equation}
As such, the SRI can be readily computed based on the temperature field prediction from the physics-based models as exemplified by Figure ~\ref{fig:data_gen}.

\begin{figure}[!t]\centering
\includegraphics[width=0.7\textwidth]{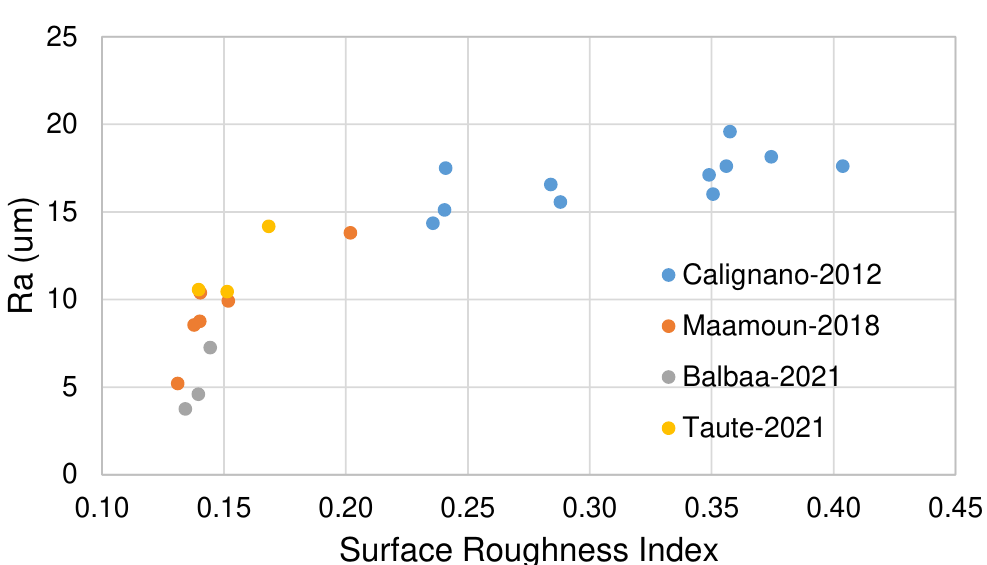}
  \caption{Surface roughness (Ra) as a function of surface roughness index (SRI).}\label{fig:sr_sri}
\end{figure}

To calculate the surface roughness with the SRI index, we conduct a literature survey on the surface roughness under various process conditions for AlSi10Mg, as listed in Table~\ref{tab:sr_data}. There are a total of 23 process conditions from 4 references, with different power $P$, scan velocity $V$, 
layer thickness $t_h$, and preheating temperature\footnote{We neglect the effects of hatch spacing as we only consider the single track simulations in the current setting.}. 
Then, we perform 23 simulations with the same process conditions as in Table~\ref{tab:sr_data} to obtain the corresponding temperature field. Subsequently, we compute the SRI based on \eqref{eq:sri} and \eqref{eq:Marangoni_force} for each case. The relationship between SRI and the measured surface roughness is shown in Figure~\ref{fig:sr_sri}. It can be verified that the SRI is an effective indicator for the surface roughness as the data from independent sources \cite{calignano2013influence,maamoun2018effect,balbaa2021role,taute2021characterization} collapses on the same trend.

\begin{table}[!h]
    \caption{Literature data collection for surface roughness modeling.}
    \label{tab:sr_data}
    \centering
    {\small \centering
    \begin{tabular}{lllllll}
    \hline
         Laser power& Scan vel. & Nominal layer & Preheating & Surface rough. & Vendor & Ref.\\
         $P$ ($W$) & $V$ ($m/s$)  & thickness $t_h$ ($\mu m$) & temp. $T_{sub}$ ($K$) & Ra ($\mu m$) &  & \\
         \hline
         120 & 0.8 &  30 & 300 & 17.61 & & \\
         190 & 0.8 & 30 & 300 & 14.35 & & \\
         120 & 0.85 & 30 & 300 & 17.11 & & \\
         120 & 0.9 & 30 & 300 & 16.01 & & \\
         155 & 0.9 & 30 & 300 & 16.57 & & \\
         190 & 0.9 & 30 & 300 & 15.11 & EOS M270 & \cite{calignano2013influence} \\
         120 & 0.95 & 30 & 300 & 17.62 & & \\
         155 & 0.95 & 30 & 300 & 15.57 & & \\
         190 & 0.95 & 30 & 300 & 17.50 & & \\
         120 & 1.0 & 30 & 300 & 19.59 & & \\
         120 & 1.25 & 30 & 300 & 18.14 & & \\
         \hline
         370 & 1.0 & 30 & 473 & 5.21 & & \\
         370 & 1.3 & 30 & 473 & 8.55 & & \\
         350 & 1.3 & 30 & 473 & 8.76 & EOS M290 & \cite{maamoun2018effect} \\
         370 & 1.5 & 30 & 473 & 10.38 & & \\
         300 & 1.3 & 30 & 473 & 9.91 & & \\
         200 & 1.3 & 30 & 473 & 13.81 & & \\
         \hline
         335 & 0.8 & 30 & 473 & 3.75 & & \\
         335 & 1.05 &  30 & 473 & 4.59 & EOS M290 & \cite{balbaa2021role} \\
         335 & 1.3 & 30 & 473 & 7.25 & & \\
         \hline
         350 & 0.921 & 50 & 300 & 14.17 &  & \\
         420 & 0.921 & 50 & 300 & 10.45 & SLM500 & \cite{taute2021characterization} \\
         490 & 0.921 & 50 & 300 & 10.57 &  & \\
         \hline
    \end{tabular}}
\vskip -0.1in
\end{table}

\begin{figure}[!t]\centering
\includegraphics[width=0.7\textwidth]{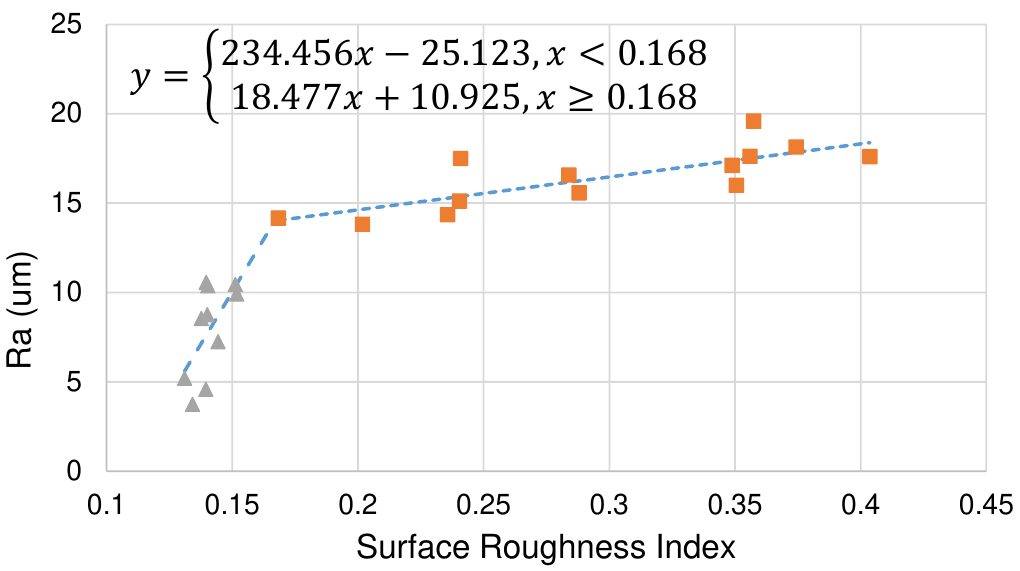}
  \caption{Bi-linear fit of surface roughness model.}\label{fig:sri_fit}
\end{figure}

To estimate and control the surface roughness in the later experiments, we calculate the SRI based on the simulated temperature field, $T(x,y,z)$, under each process condition input combination $\xib:=(P,V,T_{sub},\alpha)$, and then use this index to estimate the corresponding surface roughness with a heuristic model 
\begin{equation}
\text{Ra}=f_R(SRI[T]).
\end{equation}
Here, $f_R$ is constructed using a bi-linear fit to the relationship between SRI and surface roughness:
\begin{equation}\label{eq:ra}
f_R(s)=\left\{\begin{array}{rl}
234.456s-25.123,&\text{ if }s<0.168;\\
18.477s+10.925,&\text{ if }s\geq0.168.
\end{array}\right.
\end{equation}
In particular, the $R^2$ value for the low SRI fit (gray data points) and the high SRI fit (orange data points) are 0.61 and 0.62, respectively, indicating a reasonable linear relation (as shown in Figure~\ref{fig:sri_fit}). As a consequence, we have established the mapping between process conditions and surface roughness defects using the physics-based model described in Section~\ref{sec:physics_model} and the aforementioned surface roughness model.

\section{Neural Operator Enabled Physics Surrogates}\label{sec:no}

\subsection{Base Model: Fourier Neural Operators}

The physics-based process simulation is computationally slow, and therefore it calls for a high-fidelity machine-learning-based surrogate to facilitate online monitoring and real-time control. To this end, we employ Fourier Neural Operators (FNOs) \cite{you2022learning,li2020fourier,liu2023domain,liu2023clawno} to construct the relationship between process parameters and the resulting full-scale temperature field. FNOs adopt a convolution-based integral kernel that learns efficient solution surrogates and serves as a powerful data-driven paradigm to accelerate the PDE-solving process in scientific computing problems. The general architecture of a neural operator is comprised of a lifting layer $\mathcal{P}$ that maps the low-dimensional input function $\fb(\xb)$ into a high-dimensional latent representation $\hb(\xb)\in\real^{d_h}$, a cascade of integral operator blocks $\mathcal{J}$ in the form of the sum of a local linear operator, a nonlocal integral kernel operator, and a bias function, followed by a projection layer $\mathcal{Q}$ that projects the last hidden-layer representation to the space of the target output function. In particular, an $L$-layer neural operator can be expressed in the following form:
\begin{equation}\label{eqn:G}
\hat{\mcG}[\fb;\theta](\xb):=\mathcal{Q}\circ\mathcal{J}_{L}\circ\cdots\circ\mathcal{J}_1\circ\mathcal{P}[\fb](\xb)\approx \ub(\xb) \text{ ,}
\end{equation}
where $\tilde{\mcG}$ is the neural operator parameterized by $\theta$, acting as an efficient surrogate of the true (physical) operator $\mcG$ that maps the input function $\fb(\xb)$ to the output function $\ub(\xb)$. When employed on structured domains and uniform grids, FNO further expedites the learning process by parameterizing the integral kernel operator in the Fourier space:
\begin{align}
\mathcal{J}_{l}[\hb](\xb)=\sigma\left(W^l\hb(\xb)+ \cb^l+\mathcal{F}^{-1}[\mathcal{F}[\kappa(\cdot;\vb^l)]\cdot \mathcal{F}[\hb(\cdot)]](\xb)\right)\text{ ,}\label{eq:fno}
\end{align}
where $W^l\in\real^{d_h\times d_h}$ and $\cb^l\in\real^{d_h}$ are learnable tensors at the $l$-th layer, $\kappa\in\real^{d_h\times d_h}$ is a tensor kernel function parameterized by parameters $\vb^l$, $\mathcal{F}$ and $\mathcal{F}^{-1}$ denote the Fourier transform and its inverse, respectively, which apply the FFT algorithm to each component of $\hb$ separately.

The training process of neural operators works as the follows. Let $\{\fb_j,\ub_j\}_{j=1}^N$ be a set of observations where the input $\{\fb_j\}$ is a sequence of independent and identically distributed random fields from a known probability distribution $\mu$, and $\mcG^\dag[\fb_j](\xb)=\ub_j(\xb)$ is the ground-truth output of the map $\mcG^\dag:\fb\to\ub$. The goal is to build an approximation of $\mcG^\dag$ using the architecture of \eqref{eqn:G}, by optimizing its parameter $\theta$ in some finite-dimensional parameter space $\Theta$. In particular, $\theta$ is inferred by solving the following minimization problem
\begin{equation}\label{eqn:opt}
\min_{\theta\in\Theta}\mathbb{E}_{\fb\sim\mu}[C(\hat{\mcG}[\fb;\theta],\mcG^\dag[\fb])]\approx \min_{\theta\in\Theta}\sum_{i=1}^N[C(\mcG[\fb_i;\theta],\ub_i)],
\end{equation}
where $C$ denotes a properly defined cost functional $C:\mcU\times\mcU\rightarrow\real$. Although $\fb_i$ and $\ub_i$ are (vector) functions defined on a continuum, with the purpose of doing numerical simulations, we evaluate them on a uniform discretization of the domain defined as $\chi=\{\xb_1,\cdots,\xb_M\}\subset \omg$. With such a discretization to establish learning governing laws, a popular choice of the cost functional $C$ is the mean square error:
$$C(\hat{\mcG}[\fb_i;\theta],\ub_i):=\sum_{\xb_j\in \chi}\vertii{\mcG[\fb_i;\theta](\xb_j)-\ub_i(\xb_j)}^2.$$

Compared with the classical NNs, the most notable advantages of neural operators \cite{lu2019deeponet,li2020neural,li2020fourier,you2022nonlocal} are their resolution independence and generalizability to different input instances. Thanks to the integral architecture of \eqref{eq:fno}, the accuracy of the prediction is invariant with respect to the resolution of input parameters such as initial conditions and grids. This fact is in contrast with the classical finite-dimensional NN approaches that parameterize the mapping between finite-dimensional Euclidean spaces and hence their accuracy is tied to the resolution of the input \cite{guo2016convolutional,zhu2018bayesian,adler2017solving,bhatnagar2019prediction,khoo2021solving}. Furthermore, once trained, a neural operator is generalizable with respect to different input parameter instances; in other words, solving for a new instance of the input condition only requires a forward pass of the network. This unique property is in contrast with traditional PDE-constrained optimization techniques \cite{de2015numerical} and some other NN models that directly parameterize the solution \cite{raissi2019physics,weinan2018deep,bar2019unsupervised,smith2020eikonet,pan2020physics}. In \cite{li2020neural,li2020multipole,li2020fourier,lu2021comprehensive,you2022learning,liu2023ino}, neural operators are employed as an efficient solution surrogate for porous and/or heterogeneous materials.

\subsection{Deep Neural Operator Modeling of Melt Pool States}
The end goal of this section is to build an efficient FNO surrogate to model the melt pool states. From the formulations of FNOs, we notice that the total number of grids, $|\chi|$, as well as the size of trainable parameter $\theta$, grows exponentially with the problem dimension. Denoting $M$ as the total number of grids in $\chi$, we have $M=\textit{O}(\Delta x^{-\text{Dim}})$, where Dim denotes the dimension of the physical domain $\Omega$, and $\Delta x$ is the grid size. Then, as discussed in \cite{liu2023clawno}, the number of trainable parameters in the integral operator of FNOs grows as $\textit{O}(d_h^2 \text{mode}^{\text{Dim}})$, where $d_h$ is the dimension of layer feature at each point, $\hb(\cdot)$, and mode is the number of Fourier modes. Therefore, for the sake of computational and memory efficiency, it is desired to reduce the dimension of FNO models.

\begin{figure}[!t]\centering
\includegraphics[width=1.0\textwidth]{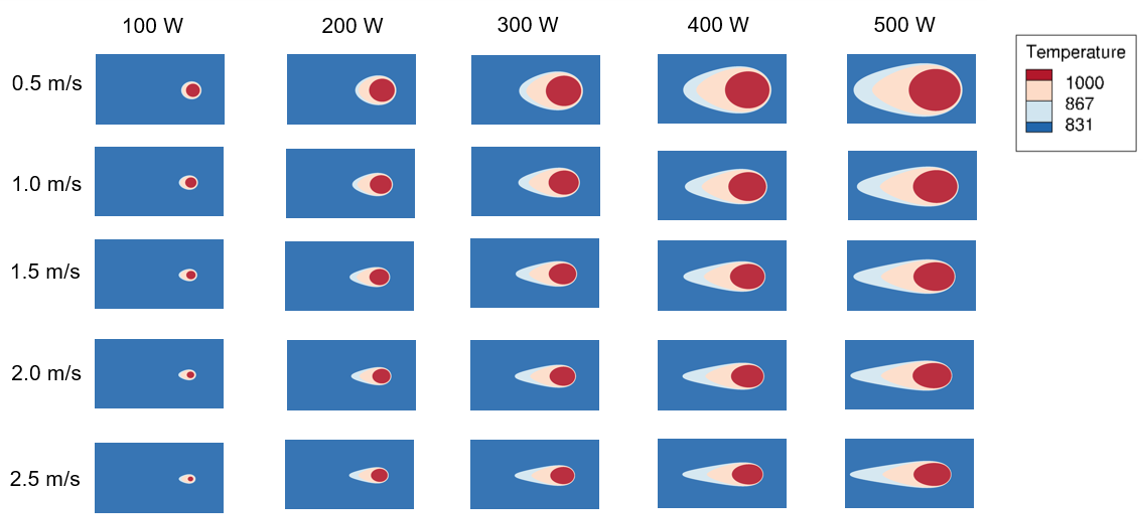}
  \caption{A schematic of the data generation process.}\label{fig:data_gen}
\end{figure}

\begin{figure}[!t]\centering
\includegraphics[width=0.76\textwidth]{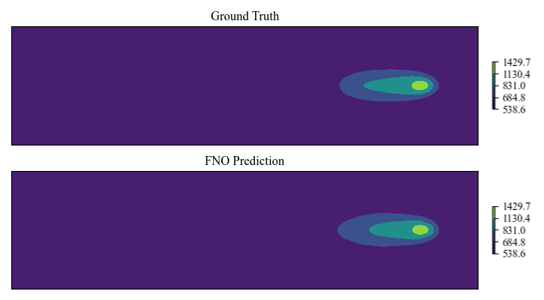}
  \caption{An example of the FNO prediction of the in-plane section.}\label{fig:z_plane_pred}
\end{figure}

\begin{figure}[!t]\centering
\includegraphics[width=0.75\textwidth]{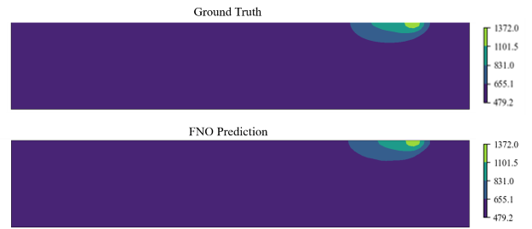}
  \caption{An example of the FNO prediction of the section along the through-thickness and scanning directions.}\label{fig:y_plane_pred}
\end{figure}

Herein, we develop a strategy to decompose the 3D temperature field in melt pool simulations to a series of 2D ones. In particular, from the generated temperature field data (see Figures \ref{fig:data_gen}-\ref{fig:y_plane_pred}), we notice that the contours of $T(x,y,z)$ are fairly smooth and almost symmetric with respect to the center line of the melt pool along the scanning direction. That means, with the 2D section cuts from the center of the melt pool in the in-plane direction (along the $x-y$ plane) and along the thickness and scanning directions (along the $x-z$ plane), one can reconstruct the full-scale temperature field. Hence, in this work we construct two 2D neural operators, to learn the temperature solution mapping of the 2D section cuts along the $x-y$ plane and the $x-z$ plane, respectively:
\begin{align}
&\hat{\mcG}_1[\xib;\theta_1](x,y)\approx T_{x-y}(x,y):=T(x,y,z)|_{z=1000},\\
&\hat{\mcG}_2[\xib;\theta_2](x,z)\approx T_{x-z}(x,z):=T(x,y,z)|_{y=0}.
\end{align}
Then, the melt pool states of \eqref{eq:sri}-\eqref{eq:Marangoni_force} can be calculated from the 2D solution surrogates:
\begin{align}
&T_{\text peak}[T]=T_{\text peak}[\hat{\mcG}_1,\hat{\mcG}_2]=\max \left[\max_{x,y}T_{x-y}(x,y),\max_{x,z}T_{x-z}(x,z)\right],\label{eq:Tpeak}\\
\nonumber&L[T]=L[\hat{\mcG}_1,\hat{\mcG}_2]\\
&=\max \left[\max_{y} \sum_{(x,y)\in \chi_{x-y}}\Delta x\mathbf{1}(T_{x-y}(x,y)-T_s),\max_{z} \sum_{(x,z)\in \chi_{x-z}}\Delta x\mathbf{1}(T_{x-z}(x,z)-T_s)\right],\label{eq:Tlength}\\
&W[T]=W[\hat{\mcG}_1,\hat{\mcG}_2]=\max_{x} \sum_{(x,y)\in \chi_{x-y}}\Delta y\mathbf{1}(T_{x-y}(x,y)-T_s),\\
&\epsilon[T]=L[T]/W[T],\\
&SRI[T]=SRI[\hat{\mcG}_1,\hat{\mcG}_2]=E \hat{L}^2\epsilon[T]^{0.25} \sqrt{\frac{2\beta}{\hat{H} \gamma_T\pi L[T](T_\text{peak}[T] - T_\text{s})}}.\label{eq:Tratio}
\end{align}
Here, $\chi_{x-y}$ and $\chi_{y-z}$ denotes the uniform discretization of the $x-y$ and $y-z$ planes, respectively. In this work, we take $\chi_{x-y}:=\{-1375,1347.5,\cdots,1375\}\times \{-220,-211.4,\cdots,220\}$ and $\chi_{x-z}:=\{-1375,1347.5,\cdots,1375\}\times \{750,760,\cdots,1000\}$, with values of $T$ on these sets by interpolating the solution of the physics-based model in Section \ref{sec:physics_modeling}.

In this context, we take laser power, scan speed, substrate temperature, and laser absorptivity as input and train two FNOs to learn the corresponding steady-state 2D temperature fields of the representative sections. With the physics-based process simulation described in Section \ref{sec:physics_modeling}, a total of 750 simulations are conducted, with 5 laser powers $[100, 200, 300, 400, 500]$ W, 5 scan speeds $[0.5, 1.0, 1.5, 2.0, 2.5]$ m/s, 5 substrate temperatures $[300, 360, 420, 480, 54]$ K, and 6 laser absorptivity $[0.1, 0.2, 0.3, 0.4, 0.5, 0.6]$, respectively. A schematic of the data generation process is shown in Figure~\ref{fig:data_gen}. Subsequently, the entire dataset, $\mcD:=\{\xib_i,T_i(x,y,z)\}_{i=1}^{N}$, is split into $N_{tr}=700$ for training and $N_{val}=50$ for testing. Then, each 3D field $T$ is interpolated onto the uniform grid sets $\chi_{x-y}$ and $\chi_{x-z}$, forming the datasets for two FNOs:
$$\mcD_{x-y}:=\{\xib_i,T_{i,x-y}(x,y)|_{(x,y)\in\chi_{x-y}}\}_{i=1}^{N},$$
$$\mcD_{x-z}:=\{\xib_i,T_{i,x-z}(x,z)|_{(x,z)\in\chi_{x-z}}\}_{i=1}^{N}.$$
$\mcD_{x-y}$ and $\mcD_{x-z}$ are used to construct $\hat{\mcG}_1$ and $\hat{\mcG}_2$, respectively. In particular, two 4-layer FNOs are adopted with a latent dimension of 20 and 12 truncated modes. The initial learning rate is set to 0.01 that decays with a factor of 0.7 every 50 epochs. The FNOs are trained for a total of 1,000 epochs using the Adam optimizer with L2 regularization of $10^{-4}$ to avoid overfitting. The relative L2-norm error is monitored and used to select the best models. The training on the in-plane section resulted in an FNO model with 0.63\% training error and 0.82\% test error, whereas the training on the section along the scan direction resulted in an FNO model with 0.66\% training error and 0.96\% test error. Examples of the trained FNO predictions are showcased in Figures~\ref{fig:z_plane_pred} and \ref{fig:y_plane_pred}. All training and test are performed on a machine with a single NVIDIA A6000 GPU.

\begin{figure}[!t]\centering
\includegraphics[width=1.0\textwidth]{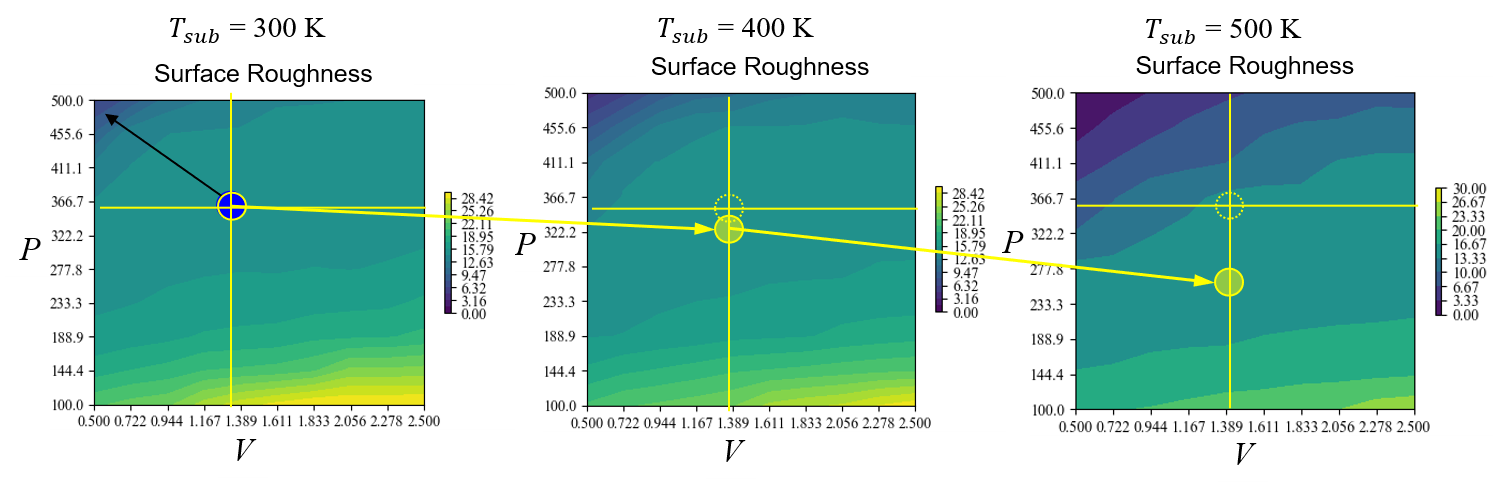}
  \caption{FNO generated process windows under different substrate temperatures.}\label{fig:process_window}
\end{figure}

Thanks to the efficiency and generalizability of FNOs, the trained models are capable of rapidly generating the process windows under different substrate temperatures. That means, given a new and unseen process condition $\xib^{\text test}=(P^{\text test},V^{\text test},T_{sub}^{\text test},\alpha^{\text test})$, the corresponding surface roughness can be evaluated via an efficient forward pass of $\hat{\mcG}_1[\xib^{\text test}]$ and $\hat{\mcG}_2[\xib^{\text test}]$, and then the numerical evaluations of \eqref{eq:sri} and \eqref{eq:Tpeak}-\eqref{eq:Tratio}. As such, we create estimations from $1000$ different combinations, and display the exemplar contour maps of the surface roughness as a function of the laser power and scan speed in Figure~\ref{fig:process_window}. Note that these simulations only take 20 seconds in total on a single GPU machine, while the full physical simulations would take 14 days on the same machine with 48 CPU cores. Taking a closer look at the 1st contour map, the blue dot on the map assumes the initial condition that one starts the simulation or the manufacturing process with. If one is intended to reduce the surface roughness, according to the contour map, one will need to increase the laser power and reduce the scan speed, moving to the upper left corner of the map in the direction of the black arrow. An interesting question these contour maps can answer is: with the increase of the substrate temperature as the printing goes on, if one is satisfied with the current surface roughness and would like to keep it constant as a nominal condition, is there a way it can be achieved? The answer is positive. According to the process window, if the scan speed is kept the same, one simply needs to reduce the laser power accordingly. On the other hand, it is obvious from the physics standpoint that if one keeps increasing the laser power and decreasing the scan speed, it will ultimately lead to keyholes. Although the current simplified physics simulation does not consider that, proper treatment is needed to avoid keyholing in the feedback control. This will be taken care of in Section~\ref{sec:integrated_dt}.

\subsection{Process Parameter Optimization}

\begin{figure}[!t]
\centering%
\begin{minipage}[b]{0.29\linewidth}
    \includegraphics[width=1.\columnwidth]{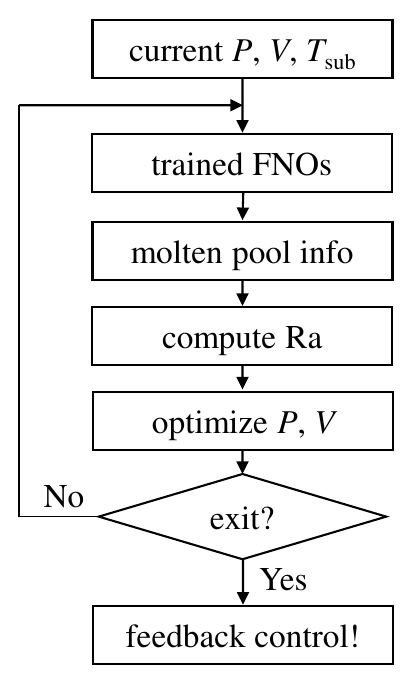}
\end{minipage}
\hspace{0.01\linewidth}
\begin{minipage}[b]{0.68\linewidth}
    \includegraphics[width=1.\columnwidth]{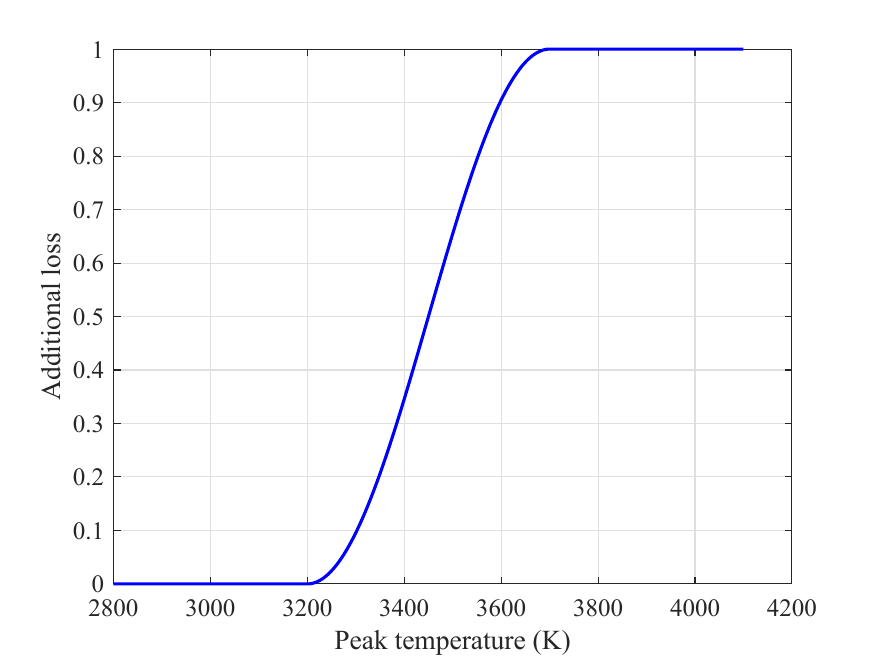}
\end{minipage}
    \caption{An illustration of the optimization workflow to search for optimal process parameters for feedback control: optimization process (left) and additional loss term for peak temperature regularization (right).}\label{fig:optimization}
\end{figure}

With a set of properly trained FNO solution surrogates at hand, we can efficiently predict the melt pool states given input scan parameters and compute the resulting defects of interest based on the relations introduced in Section~\ref{sec:defect_characterization}. Recall that the simplified simulation model does not take into account the keyholing effect. With this caveat in mind, we design an objective function that minimizes surface roughness while in the meantime regularizing the peak temperature, via penalizing the loss function when the peak temperature enters the penalization regime $[T_{\text{t1}}, T_{\text{t2}}]$. In particular, with fixed $T_{sub}$ and $\alpha$, the power $P$ and scan speed $V$ are optimized via:
\begin{align}
\nonumber P^*,V^*=&\underset{P,V}{\text{argmin}}\;f_R(SRI[\hat{\mcG}_1[P,V,T_{sub},\alpha;\theta_1],\hat{\mcG}_2[P,V,T_{sub},\alpha;\theta_2]]) \\
&+ \varphi \Phi(T_{\text peak}[\hat{\mcG}_1[P,V,T_{sub},\alpha;\theta_1],\hat{\mcG}_2[P,V,T_{sub},\alpha;\theta_2]], T_{\text{t1}}, T_{\text{t2}}).\label{eqn:optPV}
\end{align}
Here, $\Phi$ denotes a piecewise function:
\begin{displaymath}
 \Phi(T_{\text peak}, T_{\text{t1}}, T_{\text{t2}}):=\left\{\begin{array}{cc}
      0,& \text{ for }T_{\text peak}\leq T_{\text{t1}}; \\
      1,& \text{ for } T_{\text peak}\geq T_{\text{t2}}; \\
      \frac{1}{2}+\frac{1}{2} sin(\pi \cdot \frac{T_\text{peak}-T_\text{t1}}{T_\text{t2}-T_\text{t1}}-\frac{\pi}{2}),& \text{ otherwise. }
 \end{array}\right.
 \end{displaymath}
To optimize \eqref{eqn:optPV}, we freeze the trained FNOs parameters $\theta_1$ and $\theta_2$ and calculate the approximated gradient of this loss function with respect to $P$ and $V$ with the automatic differentiation engine in Pytorch. As such, we can backpropagate the loss to the input parameters and iteratively search for the optimal ones for feedback control. An illustration of this optimization workflow is displayed in Figure~\ref{fig:optimization}.

\section{Integrated Digital Twin Framework}\label{sec:integrated_dt}

\subsection{Dynamic Digital Twin}

Leveraging on the fast inference capability of the trained FNOs, the DT is able to optimize input process parameters and realize real-time closed-loop feedback control. However, the predictions thus far are still deterministic. In a real-world process, the printing quality is subject to many statistical disturbance, such as laser parameters, material properties, and machine disturbance. Therefore, the DT should be able to self-evolve based on the in-situ monitoring data, quantify the uncertain parameters, and update its prediction. Here, we present a case study of implementing such uncertainty quantification capability. We assume there is a major uncertainty in the laser absorptivity as one of the influential model parameters. To quantify the uncertainty, we collect the in-situ monitoring data of the melt pool length, which follows a probabilistic distribution, based on which we can backtrack the distribution of the input laser absorptivity. As a consequence, we can further propagate the uncertainty and predict the distribution of surface roughness. 

\begin{figure}[!t]\centering
\includegraphics[width=1.0\textwidth]{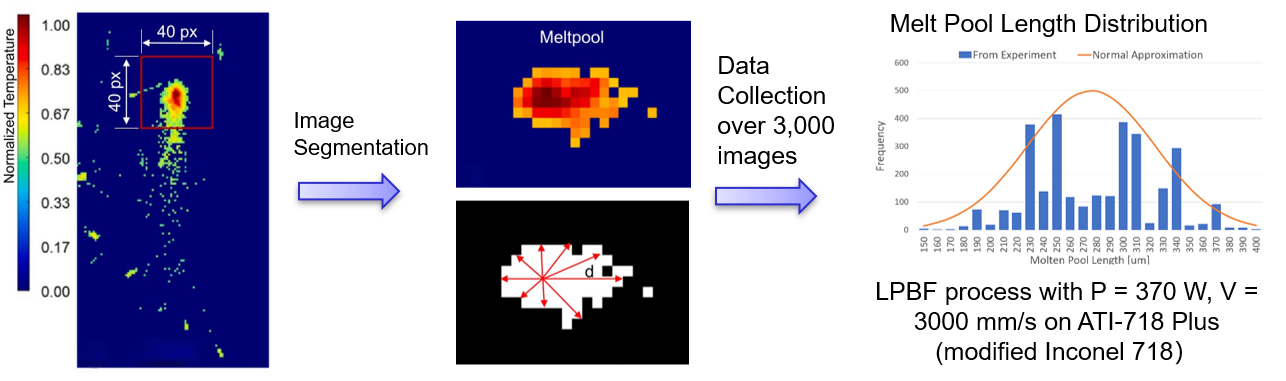}
  \caption{In-situ monitoring data processing to extract melt pool statistics.}\label{fig:mpl_distribution}
\end{figure}

To model the distribution of melt pool characteristics, we perform image segmentation to obtain the melt pool lengths from 3,000 images collected from in-situ monitoring \cite{smoqi2022monitoring}, the distribution of which is plotted in Figure~\ref{fig:mpl_distribution}. We observe that the melt pool length ranges from 150 micro-meters to 400 micro-meters. While the distribution does not follow a perfect Gaussian one, a bi-modal distribution is identified as two peaks are observed at around 240 micro-meters and 310 micro-meters. For sake of simplicity, we employed a simple normal distribution to approximate the bi-modal distribution:
\begin{equation}\label{eq:Ldistri}
L^{true}\sim \mcN[\mu^{true}_{L},(\sigma^{true}_{L})^2], \text{ where }\mu^{true}_{L}=263.30\,\mu m,\; \sigma^{true}_{L}=36.69\,\mu m.
\end{equation}
The resulting approximation is shown in Figure~\ref{fig:mpl_distribution}.

Then, we infer the laser absorptivity distribution from the melt pool image degmentation data, by matching the posterior melt pool length distribution with \eqref{eq:Ldistri}. Without loss of generality, we assume that the laser absorptivity follows a Gaussian distribution with a mean of $e_\alpha$ and a standard deviation of $\sigma_\alpha$, i.e., $\alpha\sim\mcN[\mu_\alpha,\sigma^2_\alpha]$. The overall goal is then to estimate the values of $\mu_\alpha$ and $\sigma_\alpha$ by matching the posterior melt pool distribution with the observed values. In particular, we choose the Kullback–Leibler divergence (KL divergence) as the loss function, which measures how one probability distribution is different from a second probability distribution. When both probability distributions are univariate normal distributions, the KL divergence writes:
$$D_{KL}(P||Q)=\log \dfrac{\sigma_q}{\sigma_p}+\dfrac{\sigma^2_p-\sigma^2_q+(\mu_p-\mu_q)^2}{2}.$$

To estimate the KL divergence between the estimated melt pool length and the ground truth, we sample $\alpha\sim\mcN[\mu_\alpha,\sigma^2_\alpha]$ by $S=100$ times, and then feed the resulting sample set of laser absorptivity, $\{\alpha_s\}$, to the trained FNO models to estimate the mean and standard deviation of $L[T]$:
\begin{align}
\mu_{L[T]}= &\dfrac{1}{S}\sum_{s=1}^S L[\hat{\mcG}_1[P,V,T_{sub},\alpha_s;\theta_1],\hat{\mcG}_2[P,V,T_{sub},\alpha_s;\theta_2]],\label{eq:meanL}\\
\sigma^2_{L[T]}= &\dfrac{1}{S}\sum_{s=1}^S \left(L[\hat{\mcG}_1[P,V,T_{sub},\alpha_s;\theta_1],\hat{\mcG}_2[P,V,T_{sub},\alpha_s;\theta_2]]-\mu_{L[T]}\right)^2.\label{eq:stdL}
\end{align}
In the following, we neglect the fixed parameters $P$, $V$, $T_{sub}$, $\theta_1$ and $\theta_2$, for notational simplicity. Substituting the above equations into the formulation for KL divergence, we estimate the $\mu_\alpha$ and $\sigma_\alpha$ by solving the following optimization problem:
\begin{align}
\nonumber D_{KL}(L[T]||L^{true})=&\dfrac{1}{2S}\sum_{s=1}^S \left(L[\hat{\mcG}_1[\alpha_s],\hat{\mcG}_2[\alpha_s]]-\dfrac{1}{S}\sum_{r=1}^S L[\hat{\mcG}_1[\alpha_r],\hat{\mcG}_2[\alpha_r]]\right)^2-(\sigma^{true}_L)^2\\
\nonumber &+\left(\dfrac{1}{S}\sum_{s=1}^S L[\hat{\mcG}_1[\alpha_s],\hat{\mcG}_2[\alpha_s]]-\mu^{true}_L\right)^2\\
&+\dfrac{1}{2}\log\dfrac{\dfrac{1}{S}\sum_{s=1}^S \left(L[\hat{\mcG}_1[\alpha_s],\hat{\mcG}_2[\alpha_s]]-\dfrac{1}{S}\sum_{r=1}^S L[\hat{\mcG}_1[\alpha_r],\hat{\mcG}_2[\alpha_r]]\right)^2}{(\sigma^{true}_L)^2}.\label{eq:calialpha}
\end{align}
The inferred distribution parameters of laser absorptivity are $\mu_\alpha=0.239$ and $\sigma_\alpha=0.021$. This constitutes the posterior estimation by assimilating experimental sensor data.

\subsection{Uncertainty Quantification}

With the corrected posterior laser absorptivity from in-situ data assimilation, here we further propagate the computed uncertainty in laser absorptivity to the melt pool related quantities. As of now, the FNO predicted melt pool states (e.g., the melt pool dimension, the peak temperature) are expressed as probabilistic variables. We can ultimately quantify the uncertainty in defects following the correlation in Section~\ref{sec:defect_characterization}.

\section{Virtual Demonstration}\label{sec:virtual_demo}

\subsection{Experimental Motivation}

The effect of L-PBF process parameters on surface roughness has been widely investigated in additive manufacturing research. In many of these studies, the surface roughness is evaluated based on energy density \cite{yang2019influence,wang2022part,tran2022multi,lee2022effects}. More specifically, process parameters (laser power and scan speed) are varied and correlations are drawn between energy density and surface roughness \cite{yang2019influence}. In the work by Yang et al. \cite{yang2019influence}, linear energy density is studied; however, total energy input is not evaluated (i.e heat buildup in the part during processing). It has been shown that thermal buildup during the L-PBF process influences part distortion \cite{wang2022part} and optimal process parameters are only valid for a range of interlayer temperatures \cite{tran2022multi}. The objective of this experiment is to evaluate the effect of increasing interlayer temperature on surface roughness in AlSi10Mg components fabricated via L-PBF.

\begin{figure}[!t]\centering
\includegraphics[width=1.0\textwidth]{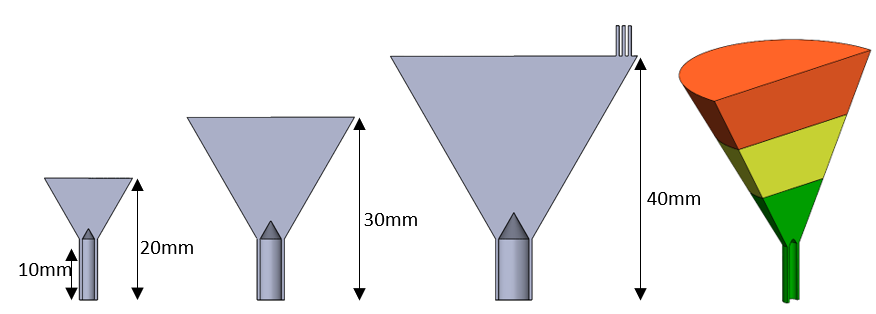}
  \caption{CAD model cross sections of the cone shaped specimens, designed to restrict heat flow to the build plate.}\label{fig:fig9}
\end{figure}

To investigate the impact of thermal buildup during the L-PBF process, cone shaped specimens with heat shunts were designed, adapted from prior work by Yavari et al. \cite{yavari2021part}. Three different height specimens (20 mm, 30 mm, and 40 mm) were designed to evaluate three initial substrate temperatures. Here, initial substrate temperature refers to the temperature of the specimen after laser scanning and powder recoating. Each specimen has a 10 mm tall heat shunt at the base to restrict heat flow to the build plate and amplify thermal build-up during the L-PBF process. Figure~\ref{fig:fig9} shows the three specimen geometries printed. Specimens are manufactured using an EOS M280 L-PBF system with EOS AlSi10Mg powder.

\begin{figure}[!t]\centering
\includegraphics[width=0.7\textwidth]{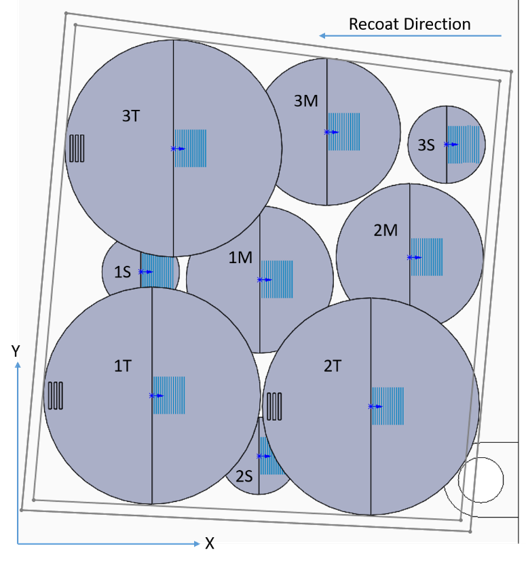}
  \caption{Specimen layout on the EOS M280 build plate. Here, the gray lines represent the thermal camera field of view.}\label{fig:fig10}
\end{figure}

The specimens were grouped into three subsets, each containing a 20 mm, 30 mm and 40 mm tall specimen. Each of these subsets is assigned a different parameter set, with laser power and scan speed being identical within a subset. By creating 3 subsets of specimens, the initial substrate temperate can be measured at three build heights, on specimens with the same process parameters. Specimens in Subset 1 were processed with a laser power of 370 W and a scan speed of 1300 mm/s, Subset 2 specimens were processed with a laser power of 296 W and a scan speed of 1300 mm/s, and Subset 3 specimens were processed with a laser power of 370 W and a scan speed of 1560 mm/s. The build layout is shown in Figure~\ref{fig:fig10}.

\begin{figure}[!t]\centering
\includegraphics[width=0.6\textwidth]{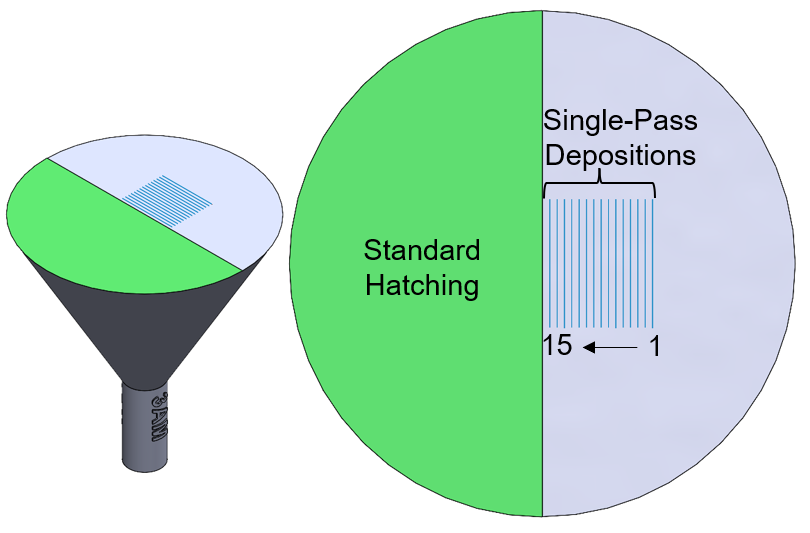}
  \caption{Single-pass depositions and standard hatches shown on the top layer of a cone specimen.}\label{fig:fig11}
\end{figure}

To evaluate the effect of substrate temperature on surface roughness, single-pass tracks and standard hatches are deposited on the top layer of each specimen. The configuration of the singles-pass depositions and standard hatches on the top layer of each cone are illustrated in Figure~\ref{fig:fig11}. Varying laser power and scan speed are selected for each of the single-pass depositions. The laser power ranges from 148 W to 370 W and the scan speed ranges from 1000 mm/s to 2080 mm/s. The laser power and scan speed parameter combinations create single pass depositions with linear energy densities ranging from 0.071 to 0.370 J/mm (cf. Table~\ref{tab:process_param}). 

\begin{table}[!h]
    \caption{Process parameters used for the single-pass depositions on the top layer of each specimen.}
    \label{tab:process_param}
    \centering
    {\small \centering
    \begin{tabular}{cccc}
    \hline
         Track \# & Laser Power (W) & Scan Speed (mm/s) & Linear Heat Input (J/mm)\\
         \hline
         1 & 370 & 1300 & 0.285\\
         2 & 296 & 1300 & 0.228\\
         3 & 370 & 1560 & 0.237\\
         4 & 370 & 1300 & 0.285\\
         5 & 370 & 2080 & 0.178\\
         6 & 296 & 1300 & 0.228\\
         7 & 222 & 1300 & 0.171\\
         8 & 370 & 1820 & 0.203\\
         9 & 370 & 1560 & 0.237\\
         10 & 370 & 1000 & 0.370\\
         11 & 148 & 1300 & 0.114\\
         12 & 370 & 1040 & 0.356\\
         13 & 370 & 1040 & 0.356\\
         14 & 148 & 2080 & 0.071\\
         15 & 296 & 1560 & 0.190\\
         \hline
    \end{tabular}}
\vskip -0.1in
\end{table}

\begin{figure}[!t]\centering
\includegraphics[width=0.6\textwidth]{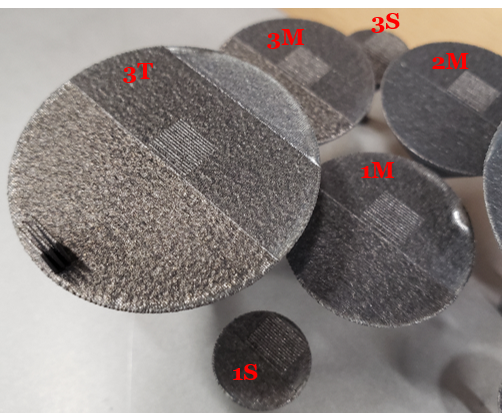}
  \caption{Top layer of the cone specimens from the completed build.}\label{fig:fig12}
\end{figure}

The manufactured specimens are shown in Figure~\ref{fig:fig12}. Of the nine specimens printed, one failed during the build. This specimen, 1T, broke off in the heat shunt region during the build. Due to the location of the 1T specimen, the build was not paused, and the remainder of the specimens were completed without interrupting the build.

\begin{figure}[!t]\centering
\includegraphics[width=0.75\textwidth]{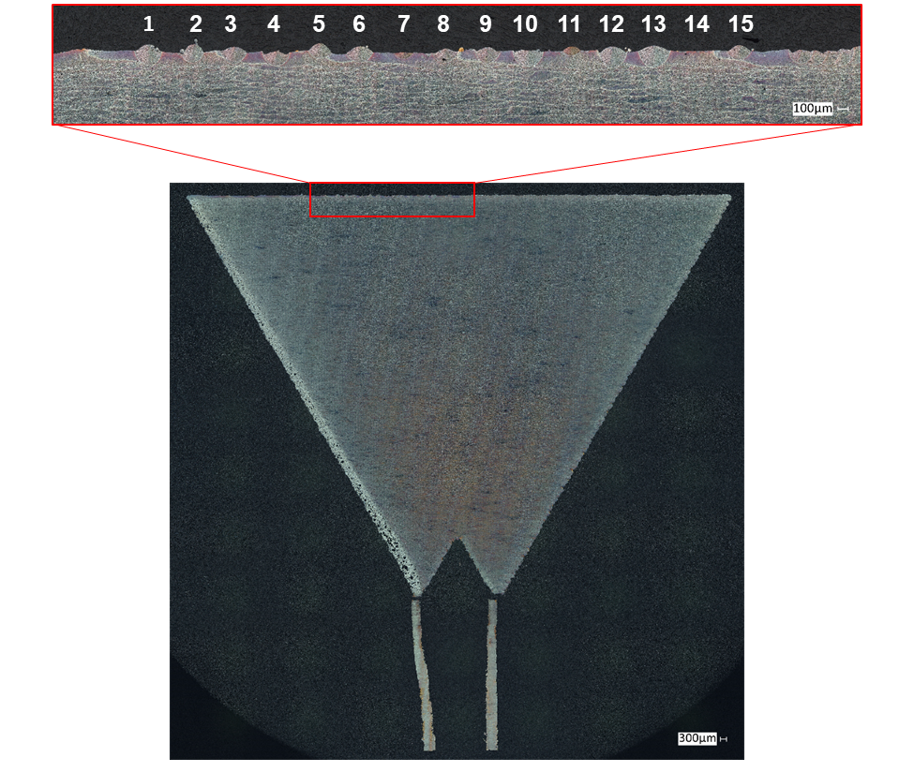}
  \caption{Cross section of specimen 3M. The highlighted section shows the 15 single tracks deposited on the top layer.}\label{fig:fig13}
\end{figure}

Specimens were sectioned, mounted in resin, and polished to reveal melt pool geometry and microstructure. Figure~\ref{fig:fig13} shows specimen 3M after being imaged using a Keyence optical microscope, and highlights the single track depositions on the top layer. From these micrographs, melt pool width, depth, and height can be measured for each single-pass deposition and the surface roughness can be evaluated on the top layer.

\begin{figure}[!t]\centering
\includegraphics[width=1.0\textwidth]{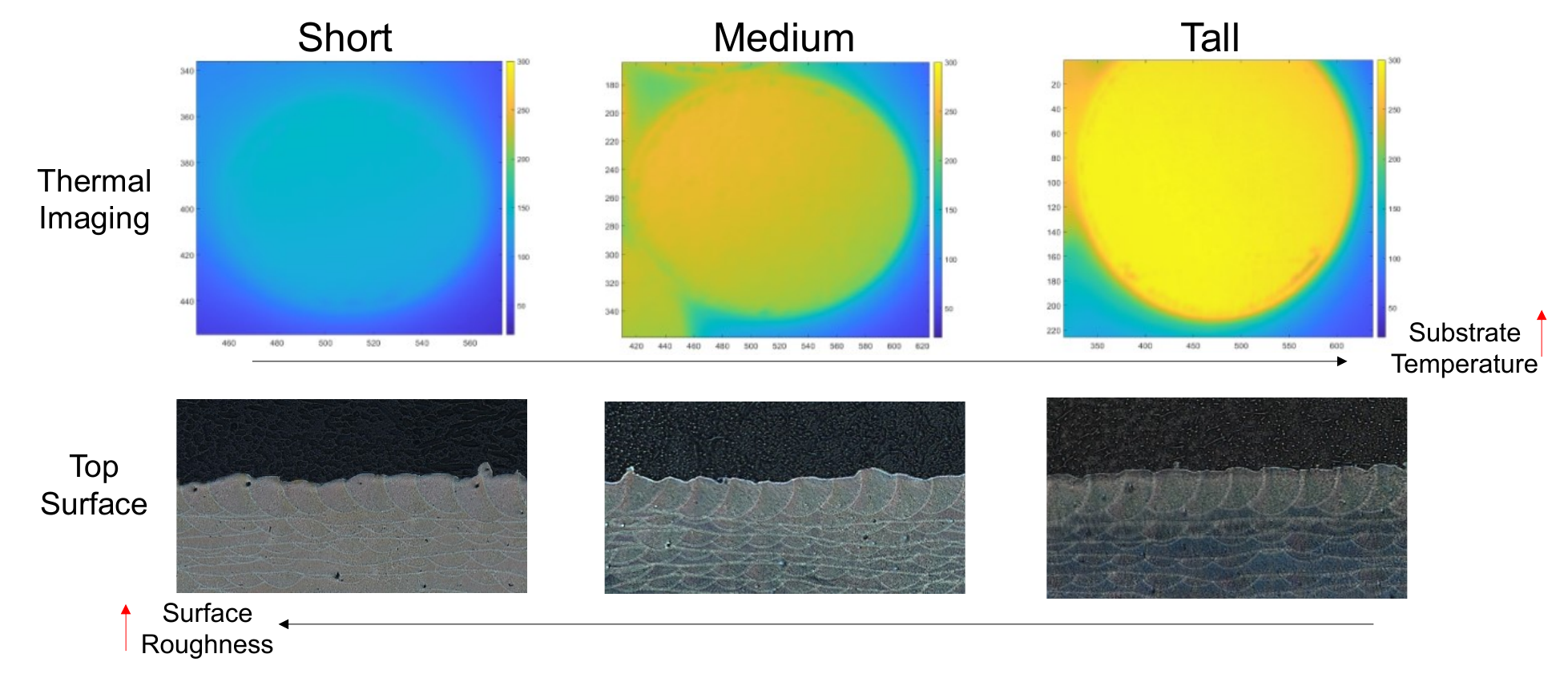}
  \caption{Thermal data collected with a FLIR X6801sc camera, showing the top layer of specimen numbers 3S, 3M, and 3T (from left to right). These images were acquired after powder recoating and before the laser scanned the top layer of the cone. Here, the color bar displays calibrated temperature in degrees Celsius. The corresponding cross sections are displayed below.}\label{fig:fig14}
\end{figure}

\begin{figure}[!t]\centering
\includegraphics[width=1.0\textwidth]{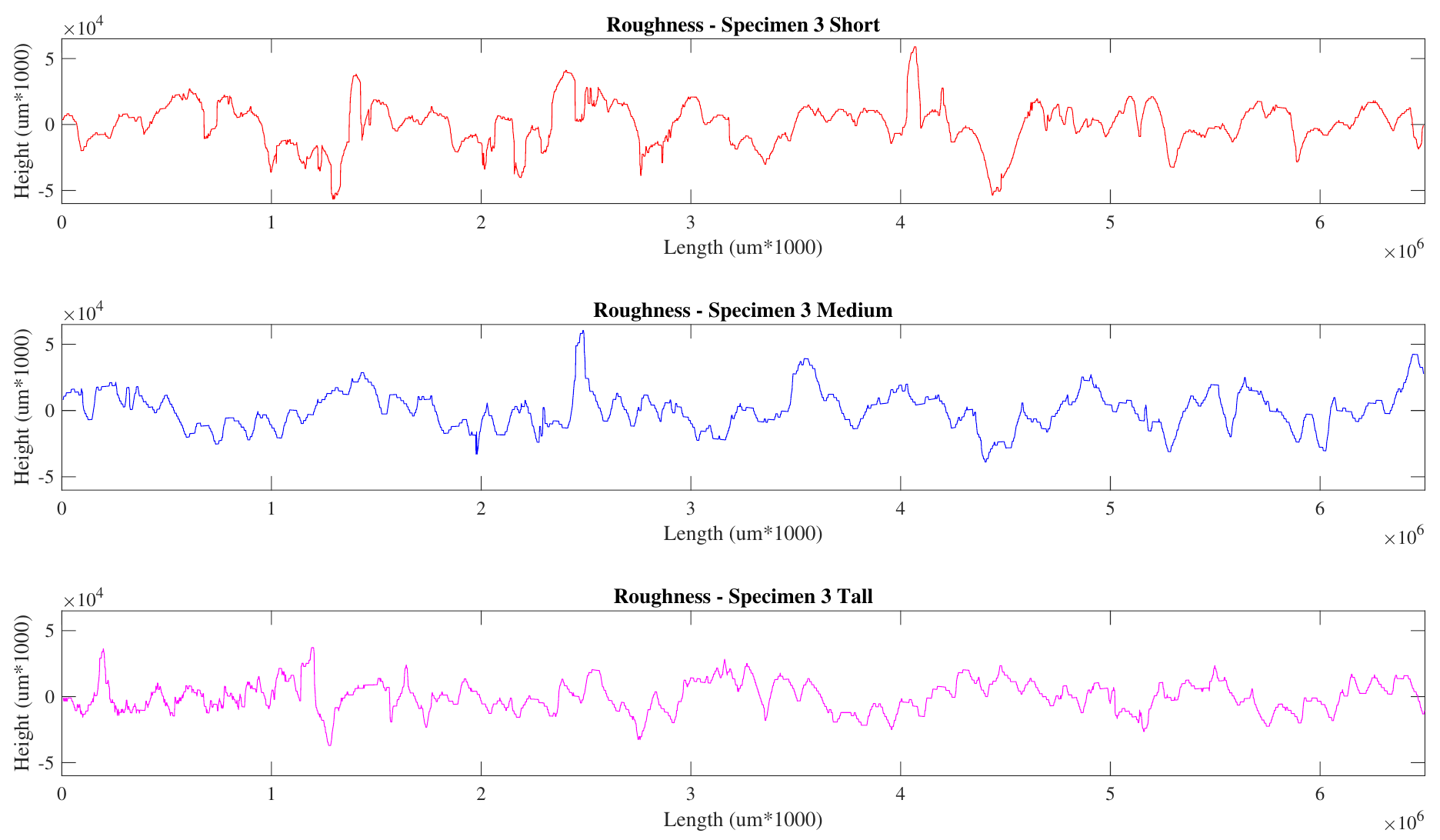}
  \caption{Illustrations of surface profiles.}\label{fig:surface_profile}
\end{figure}

To monitor the L-PBF process, multispectral and thermal data were collected. Multispectral data was collected using custom Penn State Applied Research Laboratory sensors and was acquired at 50 kHz. Additionally, thermal data was collected using a FLIR X6801sc mid-wave infrared camera at 100 frames per second. The temperature measured by the thermal camera is emissivity and surface roughness dependent; therefore, to accurately measure interlayer temperature during the build, the thermal camera was calibrated using “Method A” from Wang, et al. \cite{wang2022part}. Shown in Figure~\ref{fig:fig14} are the thermal images acquired for the top layer on specimen 3S, 3M, and 3T, with average surface temperatures of 142 °C, 235 °C, and 289 °C, respectively. The corresponding cross sections are displayed underneath the thermal images, indicating that hotter substrate results in smoother surface. A more quantitative analysis on the surface profiles can be found in Figure~\ref{fig:surface_profile}, where the average deviations from the average of the surface profile are 12.716, 11.309 and 9.278 for short, medium and tall specimens, respectively. Note that these measurements do not comply with ISO standard roughness measurements and are only used for relative comparisons.

\subsection{DT Modeling}

\begin{figure}[!t]\centering
\includegraphics[width=1.0\textwidth]{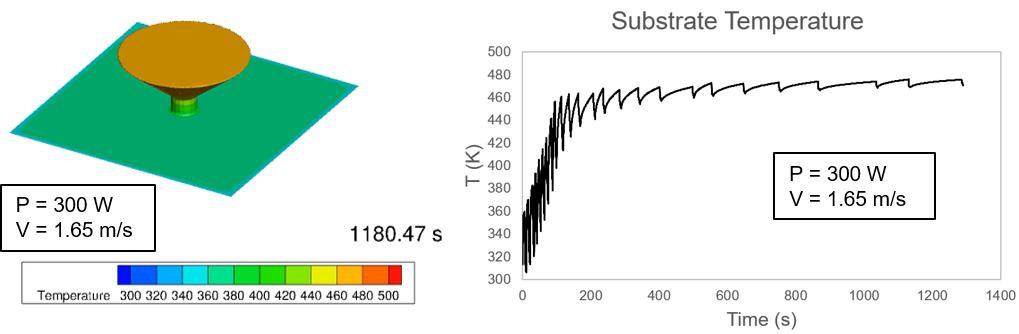}
  \caption{Part-scale virtual demonstration of DT.}\label{fig:virtual_demo}
\end{figure}

As demonstrated in Figure~\ref{fig:fig14}, a key observation from the experiment is that hotter substrate results in smoother surface, provided that keyhole is not a concern. This naturally leads to the question of whether or not one can dynamically adjust the laser power and scan speed to minimize surface roughness. In this context, we apply the developed DT framework to a part-scale cone-shaped model following the experimental setup as a virtual DT demonstration, where the heat source is homogenized and distributed to the entire layer, and 10 layers are added at a time. The initial laser power and scan speed are set to the nominal values of 300 $W$ and 1.65 $m/s$, respectively. An illustrative figure is provided in Figure~\ref{fig:virtual_demo}, along with the time history of the substrate temperature. We run two simulations of the printing process of the cone-shaped part, one without DT control and the other with DT control. The results are compared in the next section.

\section{Results and Discussion}\label{sec:discussion}

\begin{figure}[!t]
\centering%
\begin{minipage}[b]{0.485\linewidth}
    \includegraphics[width=1.\columnwidth]{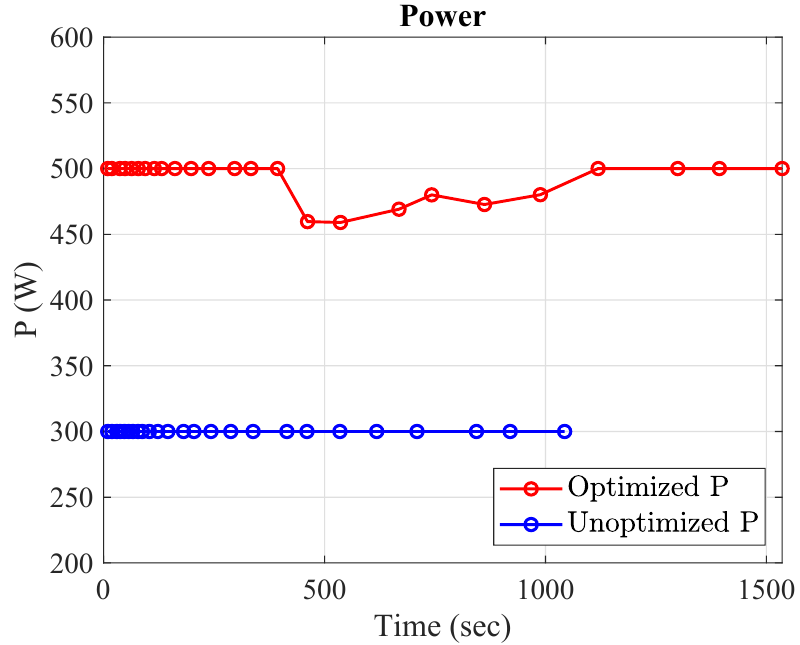}
\end{minipage}
\hspace{0.01\linewidth}
\begin{minipage}[b]{0.485\linewidth}
    \includegraphics[width=1.\columnwidth]{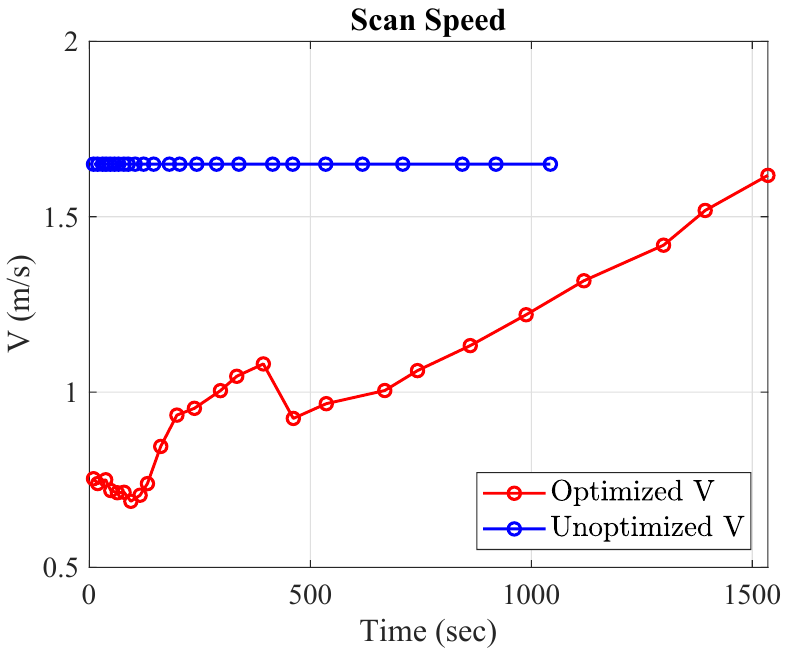}
\end{minipage}
    \caption{Comparison of process parameters with and without DT control: laser power (left) and scan speed (right).}
    \label{fig:compare_params}
\end{figure}

\begin{figure}[!t]
\centering%
\begin{minipage}[b]{0.485\linewidth}
    \includegraphics[width=1.\columnwidth]{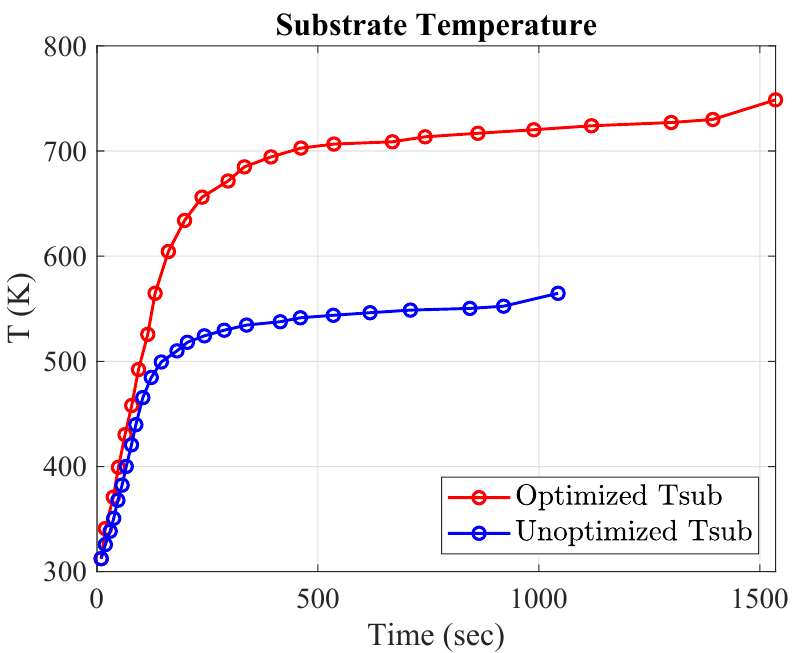}
\end{minipage}
\hspace{0.01\linewidth}
\begin{minipage}[b]{0.485\linewidth}
    \includegraphics[width=1.\columnwidth]{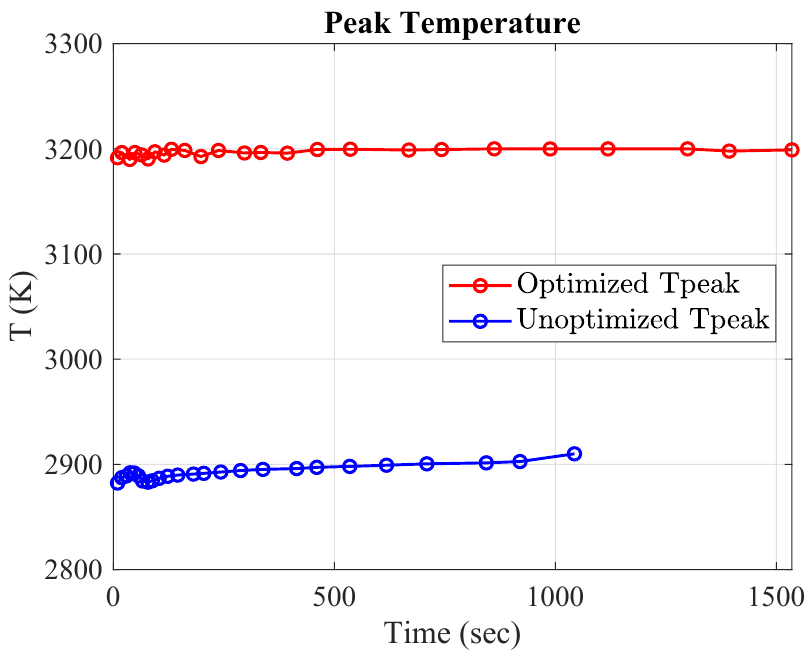}
\end{minipage}
    \caption{Comparison of substrate temperatures (left) and peak temperatures (right) with and without DT control.}
    \label{fig:compare_temps}
\end{figure}

\begin{figure}[!t]\centering
\includegraphics[width=0.6\textwidth]{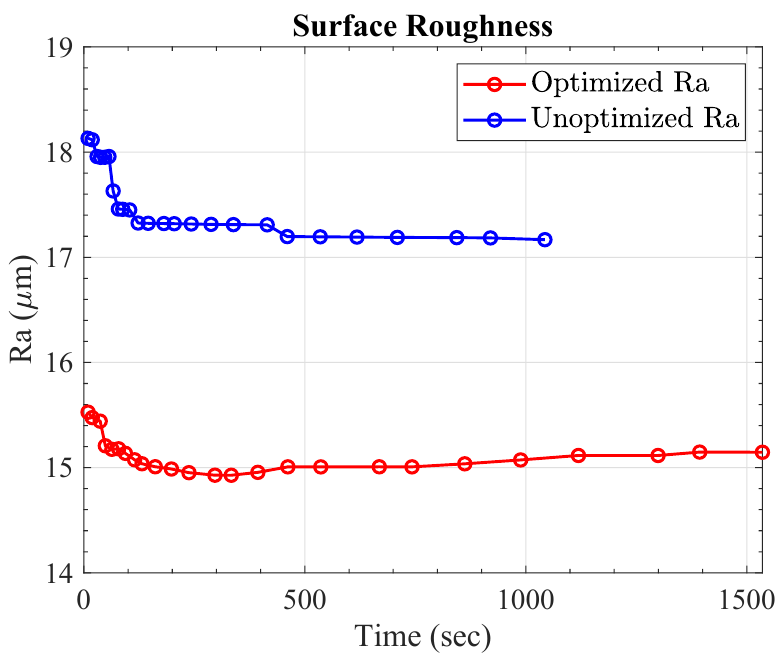}
  \caption{Comparison of the resulting surface roughness with and without DT control.}\label{fig:compare_ra}
\end{figure}

\begin{figure}[!t]
\centering%
\begin{minipage}[b]{0.495\linewidth}
    \includegraphics[width=1.\columnwidth]{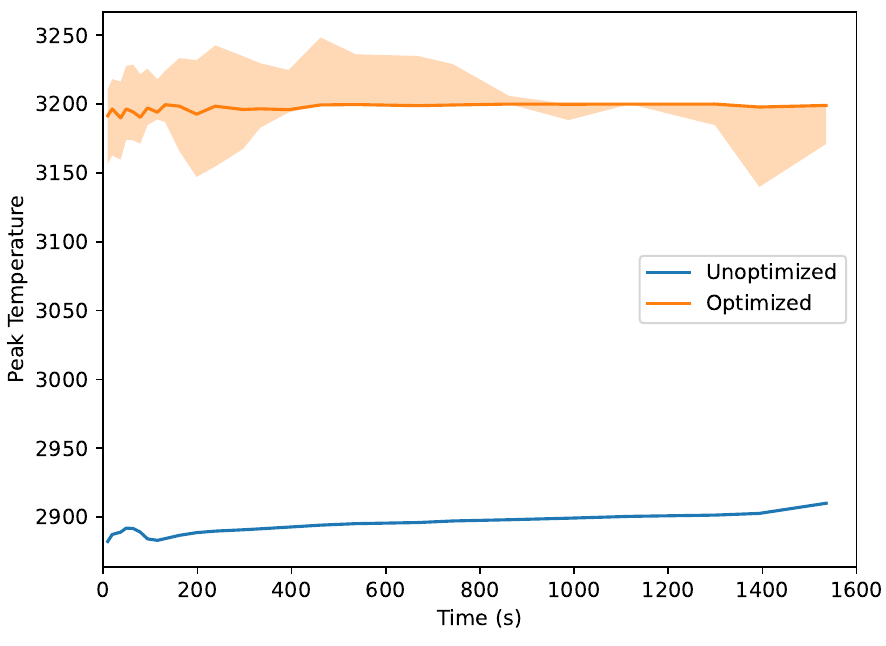}
\end{minipage}
\begin{minipage}[b]{0.495\linewidth}
    \includegraphics[width=1.\columnwidth]{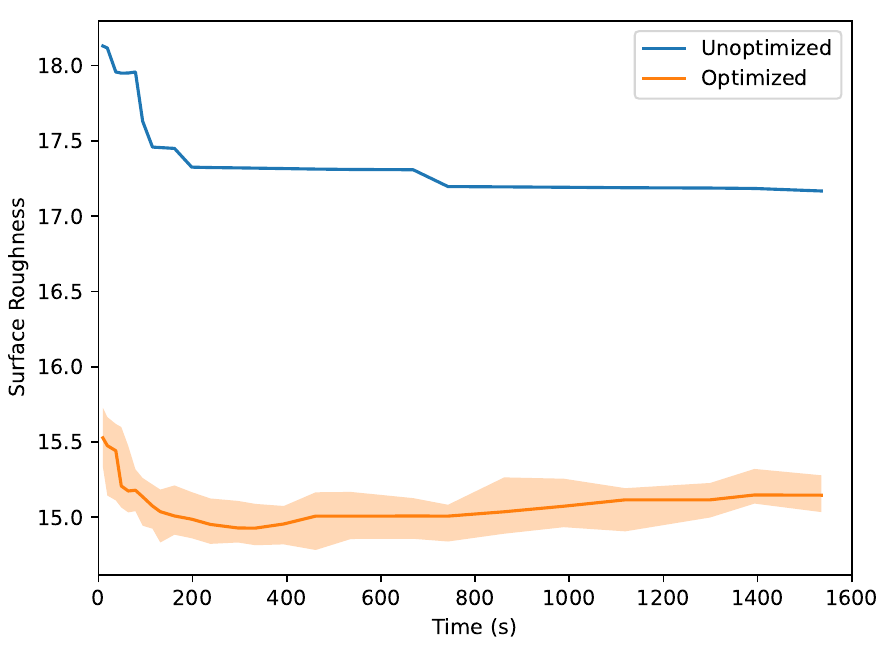}
\end{minipage}
    \caption{Uncertainty quantification of controlled peak temperature and the resulting surface roughness.}
    \label{fig:compare_uq}
\end{figure}

The comparisons between the original and DT-optimized results in terms of laser power and scan speed, substrate and melt pool peak temperatures, and the resulting surface roughness are plotted in Figures~\ref{fig:compare_params}-\ref{fig:compare_ra}, respectively. We observe that the DT dynamically increases the laser power and decreases the scan speed to slow down the scanning process. The direct consequence of this adjustment is the increase in substrate temperature, which further results in a notable decrease in surface roughness. This phenomenon is consistent with the findings in the experiment. Additionally, by looking into the peak temperature time history in Figures~\ref{fig:compare_temps} right, we observe that the DT maintains the peak temperature to be around 3,200 K, which proves the effectiveness of the peak temperature regularization scheme for keyhole mitigation. The uncertainty quantified results in terms of peak temperature and surface roughness are plotted in Figure~\ref{fig:compare_uq}.

\section{Conclusion}\label{sec:conclusion}
In this work, we developed a probabilistic DT framework for closed-loop feedback control of the L-PBF process, featuring the capability of monitoring the current state of the melt pool, updating the in-silico model, and providing online guidance for the process parameters. The proposed DT is enabled by deep neural operators that learn the hidden physics of the underlying process in the form of a function-to-function mapping from the input process parameter space to the full-scale temperature field of the melt pool. Another powerful aspect of using deep neural operators is its fast generation of process windows to aid design and manufacturing. Furthermore, an effective correlation is presented that links the extracted melt pool characteristics to the resulting defects of interest. An optimization algorithm is then designed to guide the optimal process conditions, by leveraging automatic differentiation in ML. In addition to the closed-loop feedback control, the developed DT can also assimilate in-situ monitoring data and evolve itself to closely represent the current state of the melt pool that governs the defect formation. The DT framework is also equipped with uncertainty quantification from input parameters to the final user-cared defects. 

Although the current DT control demonstration is based on synthetic data, it is envisioned that the developed DT framework can significantly improve the printing process and mitigate defects in real-world L-PBF-based metal AM processes. As a natural extension, we plan to consider the combination of synthetic data and experimental measurements, and apply our DT framework in real-time experimental designs. We also plan to explore the DT framework in other applications. Examples include experimental design of material mechanical testing \cite{you2022physics}, the non-destructive evaluation and monitoring of composites \cite{milanoski2023multi}, and so on.

\section*{Acknowledgments}
This work is supported by the U.S. Naval Air Systems Command (NAVAIR) Funding Contract No. N68335-23-C-0440. This support is gratefully acknowledged. Prahalada Rao acknowledges funding from the National Science Foundation (NSF) via Grant numbers: CMMI-2309483/1752069, OIA-1929172, PFI-TT 2322322/2044710, CMMI-1920245, ECCS-2020246, CMMI-1739696, and CMMI-1719388 for funding his research program. Understanding the causal influence of process parameters, melt pool dynamics, and thermal history on part quality was the major aspect of CMMI-2309483/1752069 CAREER award (Program Officer: Andrew Wells). Yue Yu would like to acknowledge support by the National Science Foundation under award DMS-1753031 and the AFOSR grant FA9550-22-1-0197.








\bibliography{yyu}

\begin{thebibliography}{10}
\expandafter\ifx\csname url\endcsname\relax
  \def\url#1{\texttt{#1}}\fi
\expandafter\ifx\csname urlprefix\endcsname\relax\def\urlprefix{URL }\fi
\expandafter\ifx\csname href\endcsname\relax
  \def\href#1#2{#2} \def\path#1{#1}\fi

\bibitem{frazier2014metal}
W.~E. Frazier, Metal additive manufacturing: a review, Journal of Materials
  Engineering and performance 23 (2014) 1917--1928.

\bibitem{gibson2021additive}
I.~Gibson, D.~W. Rosen, B.~Stucker, M.~Khorasani, D.~Rosen, B.~Stucker,
  M.~Khorasani, Additive manufacturing technologies, Vol.~17, Springer, 2021.

\bibitem{abdulhameed2019additive}
O.~Abdulhameed, A.~Al-Ahmari, W.~Ameen, S.~H. Mian, Additive manufacturing:
  Challenges, trends, and applications, Advances in Mechanical Engineering
  11~(2) (2019) 1687814018822880.

\bibitem{king2015laser}
W.~E. King, A.~T. Anderson, R.~M. Ferencz, N.~E. Hodge, C.~Kamath, S.~A.
  Khairallah, A.~M. Rubenchik, Laser powder bed fusion additive manufacturing
  of metals; physics, computational, and materials challenges, Applied Physics
  Reviews 2~(4).

\bibitem{yadroitsev2021fundamentals}
I.~Yadroitsev, I.~Yadroitsava, A.~Du~Plessis, E.~MacDonald, Fundamentals of
  laser powder bed fusion of metals, Elsevier, 2021.

\bibitem{tang2017prediction}
M.~Tang, P.~C. Pistorius, J.~L. Beuth, Prediction of lack-of-fusion porosity
  for powder bed fusion, Additive Manufacturing 14 (2017) 39--48.

\bibitem{promoppatum2022quantification}
P.~Promoppatum, R.~Srinivasan, S.~S. Quek, S.~Msolli, S.~Shukla, N.~S. Johan,
  S.~van~der Veen, M.~H. Jhon, Quantification and prediction of lack-of-fusion
  porosity in the high porosity regime during laser powder bed fusion of
  ti-6al-4v, Journal of Materials Processing Technology 300 (2022) 117426.

\bibitem{snyder2020understanding}
J.~C. Snyder, K.~A. Thole, Understanding laser powder bed fusion surface
  roughness, Journal of Manufacturing Science and Engineering 142~(7) (2020)
  071003.

\bibitem{fox2016effect}
J.~C. Fox, S.~P. Moylan, B.~M. Lane, Effect of process parameters on the
  surface roughness of overhanging structures in laser powder bed fusion
  additive manufacturing, Procedia Cirp 45 (2016) 131--134.

\bibitem{zhao2020critical}
C.~Zhao, N.~D. Parab, X.~Li, K.~Fezzaa, W.~Tan, A.~D. Rollett, T.~Sun, Critical
  instability at moving keyhole tip generates porosity in laser melting,
  Science 370~(6520) (2020) 1080--1086.

\bibitem{kouraytem2019effect}
N.~Kouraytem, X.~Li, R.~Cunningham, C.~Zhao, N.~Parab, T.~Sun, A.~D. Rollett,
  A.~D. Spear, W.~Tan, Effect of laser-matter interaction on molten pool flow
  and keyhole dynamics, Physical Review Applied 11~(6) (2019) 064054.

\bibitem{kapteyn2021probabilistic}
M.~G. Kapteyn, J.~V. Pretorius, K.~E. Willcox, A probabilistic graphical model
  foundation for enabling predictive digital twins at scale, Nature
  Computational Science 1~(5) (2021) 337--347.

\bibitem{gunasegaram2021towards}
D.~R. Gunasegaram, A.~Murphy, A.~Barnard, T.~DebRoy, M.~Matthews, L.~Ladani,
  D.~Gu, Towards developing multiscale-multiphysics models and their surrogates
  for digital twins of metal additive manufacturing, Additive Manufacturing 46
  (2021) 102089.

\bibitem{mukherjee2019digital}
T.~Mukherjee, T.~DebRoy, A digital twin for rapid qualification of 3d printed
  metallic components, Applied Materials Today 14 (2019) 59--65.

\bibitem{kharazmi2021data}
E.~Kharazmi, Z.~Wang, D.~Fan, S.~Rudy, T.~Sapsis, M.~S. Triantafyllou, G.~E.
  Karniadakis, From data to assessment models, demonstrated through a digital
  twin of marine risers, in: Offshore Technology Conference, OTC, 2021, p.
  D031S035R003.

\bibitem{li2021three}
E.~Li, L.~Wang, A.~Yu, Z.~Zhou, A three-phase model for simulation of heat
  transfer and melt pool behaviour in laser powder bed fusion process, Powder
  technology 381 (2021) 298--312.

\bibitem{moges2021hybrid}
T.~Moges, Z.~Yang, K.~Jones, S.~Feng, P.~Witherell, Y.~Lu, Hybrid modeling
  approach for melt-pool prediction in laser powder bed fusion additive
  manufacturing, Journal of Computing and Information Science in Engineering
  21~(5) (2021) 050902.

\bibitem{queva2020numerical}
A.~Queva, G.~Guillemot, C.~Moriconi, C.~Metton, M.~Bellet, Numerical study of
  the impact of vaporisation on melt pool dynamics in laser powder bed
  fusion-application to in718 and ti--6al--4v, Additive Manufacturing 35 (2020)
  101249.

\bibitem{khorasani2022comprehensive}
M.~Khorasani, A.~Ghasemi, M.~Leary, L.~Cordova, E.~Sharabian, E.~Farabi,
  I.~Gibson, M.~Brandt, B.~Rolfe, A comprehensive study on meltpool depth in
  laser-based powder bed fusion of inconel 718, The International Journal of
  Advanced Manufacturing Technology 120~(3) (2022) 2345--2362.

\bibitem{yeung2020meltpool}
H.~Yeung, Z.~Yang, L.~Yan, A meltpool prediction based scan strategy for powder
  bed fusion additive manufacturing, Additive Manufacturing 35 (2020) 101383.

\bibitem{li2021quantitative}
X.~Li, Q.~Guo, L.~Chen, W.~Tan, Quantitative investigation of gas flow,
  powder-gas interaction, and powder behavior under different ambient pressure
  levels in laser powder bed fusion, International Journal of Machine Tools and
  Manufacture 170 (2021) 103797.

\bibitem{heigel2015thermo}
J.~Heigel, P.~Michaleris, E.~W. Reutzel, Thermo-mechanical model development
  and validation of directed energy deposition additive manufacturing of
  ti--6al--4v, Additive manufacturing 5 (2015) 9--19.

\bibitem{strano2013surface}
G.~Strano, L.~Hao, R.~M. Everson, K.~E. Evans, Surface roughness analysis,
  modelling and prediction in selective laser melting, Journal of Materials
  Processing Technology 213~(4) (2013) 589--597.

\bibitem{liu2023review}
J.~Liu, J.~Ye, D.~Silva~Izquierdo, A.~Vinel, N.~Shamsaei, S.~Shao, A review of
  machine learning techniques for process and performance optimization in laser
  beam powder bed fusion additive manufacturing, Journal of Intelligent
  Manufacturing 34~(8) (2023) 3249--3275.

\bibitem{sing2021perspectives}
S.~L. Sing, C.~Kuo, C.~Shih, C.~Ho, C.~K. Chua, Perspectives of using machine
  learning in laser powder bed fusion for metal additive manufacturing, Virtual
  and Physical Prototyping 16~(3) (2021) 372--386.

\bibitem{okaro2019automatic}
I.~A. Okaro, S.~Jayasinghe, C.~Sutcliffe, K.~Black, P.~Paoletti, P.~L. Green,
  Automatic fault detection for laser powder-bed fusion using semi-supervised
  machine learning, Additive Manufacturing 27 (2019) 42--53.

\bibitem{du2021physics}
Y.~Du, T.~Mukherjee, T.~DebRoy, Physics-informed machine learning and
  mechanistic modeling of additive manufacturing to reduce defects, Applied
  Materials Today 24 (2021) 101123.

\bibitem{wang2022process}
P.~Wang, Y.~Yang, N.~S. Moghaddam, Process modeling in laser powder bed fusion
  towards defect detection and quality control via machine learning: The
  state-of-the-art and research challenges, Journal of Manufacturing Processes
  73 (2022) 961--984.

\bibitem{liao2024deep}
S.~Liao, T.~Xue, J.~Cao, Deep learning based reconstruction of transient 3d
  melt pool geometries in laser powder bed fusion from coaxial melt pool
  images, Manufacturing Letters.

\bibitem{larsen2022deep}
S.~Larsen, P.~A. Hooper, Deep semi-supervised learning of dynamics for anomaly
  detection in laser powder bed fusion, Journal of Intelligent Manufacturing
  33~(2) (2022) 457--471.

\bibitem{tan2019encoder}
Y.~Tan, B.~Jin, A.~Nettekoven, Y.~Chen, Y.~Yue, U.~Topcu,
  A.~Sangiovanni-Vincentelli, An encoder-decoder based approach for anomaly
  detection with application in additive manufacturing, in: 2019 18th IEEE
  international conference on machine learning and applications (ICMLA), IEEE,
  2019, pp. 1008--1015.

\bibitem{li2023statistical}
Y.~Li, S.~Mojumder, Y.~Lu, A.~A. Amin, J.~Guo, X.~Xie, W.~Chen, G.~J. Wagner,
  J.~Cao, W.~K. Liu, Statistical parameterized physics-based machine learning
  digital twin models for laser powder bed fusion process, arXiv preprint
  arXiv:2311.07821.

\bibitem{you2022learning}
H.~You, Q.~Zhang, C.~J. Ross, C.-H. Lee, Y.~Yu, {Learning Deep Implicit Fourier
  Neural Operators (IFNOs) with Applications to Heterogeneous Material
  Modeling}, arXiv preprint arXiv:2203.08205.

\bibitem{li2020fourier}
Z.~Li, N.~B. Kovachki, K.~Azizzadenesheli, K.~Bhattacharya, A.~Stuart,
  A.~Anandkumar, et~al., Fourier neural operator for parametric partial
  differential equations, in: International Conference on Learning
  Representations, 2020.

\bibitem{liu2023domain}
N.~Liu, S.~Jafarzadeh, Y.~Yu, Domain agnostic fourier neural operators, arXiv
  preprint arXiv:2305.00478.

\bibitem{liu2023clawno}
N.~Liu, Y.~Fan, X.~Zeng, M.~Kl{\"o}wer, Y.~Yu, Harnessing the power of neural
  operators with automatically encoded conservation laws, arXiv preprint
  arXiv:2312.11176.

\bibitem{jafarzadeh2024peridynamic}
S.~Jafarzadeh, S.~Silling, N.~Liu, Z.~Zhang, Y.~Yu, Peridynamic neural
  operators: A data-driven nonlocal constitutive model for complex material
  responses, arXiv preprint arXiv:2401.06070.

\bibitem{liu2023ino}
N.~Liu, Y.~Yu, H.~You, N.~Tatikola, Ino: Invariant neural operators for
  learning complex physical systems with momentum conservation, in:
  International Conference on Artificial Intelligence and Statistics, PMLR,
  2023, pp. 6822--6838.

\bibitem{piasecka2022experimental}
M.~Piasecka, A.~Piasecki, N.~Dadas, Experimental study and cfd modeling of
  fluid flow and heat transfer characteristics in a mini-channel heat sink
  using simcenter star-ccm+ software, Energies 15~(2) (2022) 536.

\bibitem{du2022high}
Y.~Du, T.~Mukherjee, N.~Finch, A.~De, T.~DebRoy, High-throughput screening of
  surface roughness during additive manufacturing, Journal of Manufacturing
  Processes 81 (2022) 65--77.

\bibitem{calignano2013influence}
F.~Calignano, D.~Manfredi, E.~Ambrosio, L.~Iuliano, P.~Fino, Influence of
  process parameters on surface roughness of aluminum parts produced by dmls,
  The International Journal of Advanced Manufacturing Technology 67 (2013)
  2743--2751.

\bibitem{maamoun2018effect}
A.~H. Maamoun, Y.~F. Xue, M.~A. Elbestawi, S.~C. Veldhuis, Effect of selective
  laser melting process parameters on the quality of al alloy parts: Powder
  characterization, density, surface roughness, and dimensional accuracy,
  Materials 11~(12) (2018) 2343.

\bibitem{balbaa2021role}
M.~Balbaa, A.~Ghasemi, E.~Fereiduni, M.~Elbestawi, S.~Jadhav, J.-P. Kruth, Role
  of powder particle size on laser powder bed fusion processability of alsi10mg
  alloy, Additive manufacturing 37 (2021) 101630.

\bibitem{taute2021characterization}
C.~Taute, H.~M{\"o}ller, A.~Du~Plessis, M.~Tshibalanganda, M.~Leary,
  Characterization of additively manufactured alsilomg cubes with different
  porosities, Journal of the Southern African Institute of Mining and
  Metallurgy 121~(4) (2021) 143--150.

\bibitem{lu2019deeponet}
L.~Lu, P.~Jin, G.~E. Karniadakis, Deeponet: Learning nonlinear operators for
  identifying differential equations based on the universal approximation
  theorem of operators, arXiv preprint arXiv:1910.03193.

\bibitem{li2020neural}
Z.~Li, N.~Kovachki, K.~Azizzadenesheli, B.~Liu, K.~Bhattacharya, A.~Stuart,
  A.~Anandkumar, Neural operator: Graph kernel network for partial differential
  equations, arXiv preprint arXiv:2003.03485.

\bibitem{you2022nonlocal}
H.~You, Y.~Yu, M.~D'Elia, T.~Gao, S.~Silling, {Nonlocal kernel network (NKN): a
  stable and resolution-independent deep neural network}, arXiv preprint
  arXiv:2201.02217.

\bibitem{guo2016convolutional}
X.~Guo, W.~Li, F.~Iorio, Convolutional neural networks for steady flow
  approximation, in: Proceedings of the 22nd ACM SIGKDD international
  conference on knowledge discovery and data mining, 2016, pp. 481--490.

\bibitem{zhu2018bayesian}
Y.~Zhu, N.~Zabaras, Bayesian deep convolutional encoder--decoder networks for
  surrogate modeling and uncertainty quantification, Journal of Computational
  Physics 366 (2018) 415--447.

\bibitem{adler2017solving}
J.~Adler, O.~{\"O}ktem, Solving ill-posed inverse problems using iterative deep
  neural networks, Inverse Problems 33~(12) (2017) 124007.

\bibitem{bhatnagar2019prediction}
S.~Bhatnagar, Y.~Afshar, S.~Pan, K.~Duraisamy, S.~Kaushik, Prediction of
  aerodynamic flow fields using convolutional neural networks, Computational
  Mechanics 64~(2) (2019) 525--545.

\bibitem{khoo2021solving}
Y.~Khoo, J.~Lu, L.~Ying, Solving parametric pde problems with artificial neural
  networks, European Journal of Applied Mathematics 32~(3) (2021) 421--435.

\bibitem{de2015numerical}
J.~C. De~los Reyes, Numerical PDE-constrained optimization, Springer, 2015.

\bibitem{raissi2019physics}
M.~Raissi, P.~Perdikaris, G.~E. Karniadakis, Physics-informed neural networks:
  A deep learning framework for solving forward and inverse problems involving
  nonlinear partial differential equations, Journal of Computational Physics
  378 (2019) 686--707.

\bibitem{weinan2018deep}
E.~Weinan, B.~Yu, The deep ritz method: A deep learning-based numerical
  algorithm for solving variational problems, Communications in Mathematics and
  Statistics 6~(1).

\bibitem{bar2019unsupervised}
L.~Bar, N.~Sochen, Unsupervised deep learning algorithm for pde-based forward
  and inverse problems, arXiv preprint arXiv:1904.05417.

\bibitem{smith2020eikonet}
J.~D. Smith, K.~Azizzadenesheli, Z.~E. Ross, Eikonet: Solving the eikonal
  equation with deep neural networks, IEEE Transactions on Geoscience and
  Remote Sensing.

\bibitem{pan2020physics}
S.~Pan, K.~Duraisamy, Physics-informed probabilistic learning of linear
  embeddings of nonlinear dynamics with guaranteed stability, SIAM Journal on
  Applied Dynamical Systems 19~(1) (2020) 480--509.

\bibitem{li2020multipole}
Z.~Li, N.~Kovachki, K.~Azizzadenesheli, B.~Liu, A.~Stuart, K.~Bhattacharya,
  A.~Anandkumar, Multipole graph neural operator for parametric partial
  differential equations, Advances in Neural Information Processing Systems 33.

\bibitem{lu2021comprehensive}
L.~Lu, X.~Meng, S.~Cai, Z.~Mao, S.~Goswami, Z.~Zhang, G.~E. Karniadakis, A
  comprehensive and fair comparison of two neural operators (with practical
  extensions) based on fair data, arXiv preprint arXiv:2111.05512.

\bibitem{smoqi2022monitoring}
Z.~Smoqi, A.~Gaikwad, B.~Bevans, M.~H. Kobir, J.~Craig, A.~Abul-Haj,
  A.~Peralta, P.~Rao, Monitoring and prediction of porosity in laser powder bed
  fusion using physics-informed meltpool signatures and machine learning,
  Journal of Materials Processing Technology 304 (2022) 117550.

\bibitem{yang2019influence}
T.~Yang, T.~Liu, W.~Liao, E.~MacDonald, H.~Wei, X.~Chen, L.~Jiang, The
  influence of process parameters on vertical surface roughness of the alsi10mg
  parts fabricated by selective laser melting, Journal of Materials Processing
  Technology 266 (2019) 26--36.

\bibitem{wang2022part}
Q.~Wang, P.~Michaleris, M.~Pantano, C.~Li, Y.~Ren, A.~R. Nassar, Part-scale
  thermal evolution and post-process distortion of inconel-718 builds
  fabricated by laser powder bed fusion, Journal of Manufacturing Processes 81
  (2022) 865--880.

\bibitem{tran2022multi}
H.-C. Tran, Y.-L. Lo, T.-N. Le, A.~K.-T. Lau, H.-Y. Lin, Multi-scale simulation
  approach for identifying optimal parameters for fabrication ofhigh-density
  inconel 718 parts using selective laser melting, Rapid Prototyping Journal
  28~(1) (2022) 109--125.

\bibitem{lee2022effects}
H.-J. Lee, Effects of the energy density on pores, hardness, surface roughness,
  and tensile characteristics of deposited astm 316l specimens with powder-bed
  fusion process, Materials 15~(19) (2022) 6672.

\bibitem{yavari2021part}
R.~Yavari, Z.~Smoqi, A.~Riensche, B.~Bevans, H.~Kobir, H.~Mendoza, H.~Song,
  K.~Cole, P.~Rao, Part-scale thermal simulation of laser powder bed fusion
  using graph theory: Effect of thermal history on porosity, microstructure
  evolution, and recoater crash, Materials \& Design 204 (2021) 109685.

\bibitem{you2022physics}
H.~You, Q.~Zhang, C.~J. Ross, C.-H. Lee, M.-C. Hsu, Y.~Yu, A physics-guided
  neural operator learning approach to model biological tissues from digital
  image correlation measurements, Journal of Biomechanical Engineering 144~(12)
  (2022) 121012.

\bibitem{milanoski2023multi}
D.~Milanoski, G.~Galanopoulos, D.~Zarouchas, T.~Loutas, Multi-level damage
  diagnosis on stiffened composite panels based on a damage-uninformative
  digital twin, Structural Health Monitoring 22~(2) (2023) 1437--1459.

\end{thebibliography}
\bibliographystyle{elsarticle-num}

\medskip
Received xxxx 20xx; revised xxxx 20xx; early access xxxx 20xx.
\medskip

\end{document}